\documentclass[11pt]{article}
\usepackage{latexsym}
 \csname
@addtoreset\endcsname{equation}{section}

\def\bec{\begin{center}}
\def\ec{\end{center}}
\def\a{\alpha}  

\def\b{\beta}   
\def\c{\gamma} 
\def\C{\Gamma}
\def\d{\delta} 

\def\e{\epsilon} 

\def\f{\phi}
\def\F{\Phi}

\def\k{\kappa}
\def\l{\lambda}
\def\L{\Lambda}
\def\m{\mu}
\def\n{\nu}

\def\t{\tau}

\def\o{\omega}

\def\cL{{\cal L}}

\def\cF{{\cal F}}

\def\cM{{\cal M}}

\def\cR{{\cal R}}

\let\la=\label

\def\nn{\nonumber}
\newcommand{\eq}[1]{(\ref{#1})}

\newcommand{\w}[1]{\\[0.#1cm]}
\def\be{\begin{equation}}
\def\ee{\end{equation}}
\def\bea{\begin{eqnarray}}
\def\eea{\end{eqnarray}}
\def\ba{\begin{array}}
\def\ea{\end{array}}

\def\mx#1#2#3#4{\left#1\begin{array}{#2} #3 \end{array}\right#4}

\def\ft#1#2{{\textstyle{{\scriptstyle #1}
\over {\scriptstyle #2}}}}

\def\ket#1{|#1\rangle}

\begin{document}
\begin{flushright}
ROM2F-04/36\\
MIFP-04-25\\
UUITP-29/04 \\
hep-th/yymmddd \vskip 8pt {\today}
\end{flushright}

\vspace{10pt}

\begin{center}

%%%%%%%%%%%%%%%%%%%%%%%%%%%%%%%%%%%%%%%%%%%%%%%%%%%%%%%%%%%%%%%%%%%%

{\Large\sc On Higher Spins with a Strong $Sp(2,R)$ Condition}

%%%%%%%%%%%%%%%%%%%%%%%%%%%%%%%%%%%%%%%%%%%%%%%%%%%%%%%%%%%%%%%%%%%%

\vspace{20pt}
{\sc A. Sagnotti${}^{1}$, E. Sezgin$^{2}$ and P. Sundell$^{3}$}\\[15pt]

{${}^1$\it\small Dipartimento di Fisica, Universit\`a di Roma
``Tor Vergata''\\ I.N.F.N., Sezione di Roma ``Tor Vergata'' \\ Via
della Ricerca Scientifica 1, 00133 Roma, ITALY }\vspace{5pt}

{${}^2$\it\small George P. and Cynthia W. Mitchell Institute for
Fundamental
 Physics\\
Texas A\&M University, College Station, TX 77843-4242,
USA}\vspace{5pt}

{${}^3$\it\small Department for Theoretical Physics, Uppsala
Universitet\\ Box 803, 751 08 Uppsala, SWEDEN}

%%%%%%%%%%%%%%%%%%%%%%%%%%%%%%%%%%%%%%%%%%%%%%%%%%%%%%%%%%%%%%%%%%%%

\vspace{15pt} {\sc\large Abstract}\end{center}

We report on an analysis of the Vasiliev construction for minimal
bosonic higher-spin master fields with oscillators that are
vectors of $SO(D-1,2)$ and doublets of $Sp(2,R)$. We show that, if
the original master field equations are supplemented with a strong
$Sp(2,R)$ projection of the 0-form while letting the 1-form adjust
to the resulting Weyl curvatures, the linearized on-shell
constraints exhibit both the proper mass terms and a geometric
gauge symmetry with unconstrained, traceful parameters. We also
address some of the subtleties related to the strong projection
and the prospects for obtaining a finite curvature expansion.

\setcounter{page}{1}

\vskip 36pt {\sl Based in part on the lectures presented by
P.~Sundell at the ``First Solvay Workshop on Higher-Spin Gauge
Theory'', Brussels, May 12-14 \ 2004, on the lectures presented by
A.~Sagnotti at ``Modern Trends in String Theory, II'', Oporto,
June 21-26 \ 2004, on the talk presented by A.~Sagnotti at the
``37th Symposium Ahrenshoop'', Berlin-Schmoeckwitz, August 23-27 \
2004, and on the talks presented by E.~Sezgin at the ``Third
International Symposium on Quantum Theory and Symmetries'',
(QTS3), Cincinnati, Ohio, USA, September 10 - 14, 2003 and at
``Deserfest \& The Status of M-Theory'', Ann Arbor, Michigan, USA,
3-7 April 2004}

\pagebreak

\tableofcontents

\pagebreak

%%%%%%%%%%%%%%%%%%%%%%%%%%%%%%%%%%%%%%%%%%%%%%%%%%%%%%%%%%%%%%%%%%%%%

\section{INTRODUCTION}

%%%%%%%%%%%%%%%%%%%%%%%%%%%%%%%%%%%%%%%%%%%%%%%%%%%%%%%%%%%%%%%%%%%%%

A consistent interacting higher-spin (HS) gauge theory can be
viewed as a generalization of Einstein gravity by the inclusion of
infinitely many massless HS fields, together with finitely many
lower-spin fields including one or more scalars \cite{v1}. There
are many reasons, if not a rigorous proof, that point to the
inevitable presence of infinite towers of fields in these systems.

Although HS gauge theories were studied for a long time for their
own sake \footnote{A notable exception is \cite{bsst}, where a
connection between massless HS gauge theories and the
eleven-dimensional supermembrane was proposed.} (see, for
instance, \cite{Vasiliev:1999ba,5d,Bouatta:2004kk} for reviews and
more references to the original literature), the well-known recent
advances on holography in AdS have shifted to some extent the
emphasis toward seeking connections with String/M Theory. Indeed,
tensionless strings should lie behind massless higher spins in
AdS, as proposed in \cite{Sundborg:2000wp}, with the leading Regge
trajectory becoming massless at some critical value of the tension
\cite{Sezgin:2002rt}, and interesting work in support of the
latter proposal was recently done in \cite{Bianchi:2003wx}. The
theory of massless HS fields has a long history, and the
literature has been growing considerably in recent years.
Therefore, rather than attempting to review the subject, we shall
stress that consistent, fully interacting \emph{and}
supersymmetric HS gauge theories have not been constructed yet in
dimensions beyond four (although the linearized field equations
\cite{Sezgin:2001yf} and certain cubic couplings
\cite{Alkalaev:2002rq} are known in 5D, and to a lesser extent in
7D \cite{Sezgin:2002rt}). This was also the case for bosonic HS
gauge theories, until Vasiliev \cite{vd} recently proposed a set
of nonlinear field equations for totally symmetric tensors in
arbitrary dimensions. The new ingredient is the use of an
$Sp(2,R)$ doublet of oscillators valued in the vector
representation of $SO(D-1,2)$ \cite{Bars:2001um}, called $Y^{iA}$
in the following, as opposed to the bosonic spinor oscillators
\cite{gunaydin} used in the original construction of the 4D theory
\cite{v1}.

The two formulations\footnote{The precise relation between the
vector and spinor-oscillator formulations of the Vasiliev
equations is yet to be determined.} rest on the same sequence of
minimal HS Lie algebra extensions of $SO(D-1,2)$ \cite{misha,ps}.
Not surprisingly, the \emph{associative} nature of the oscillator
$\star$-product algebra, discussed in Section 3.2, is reflected in
the possibility of extending the HS field equations to
matrix-valued master fields associated to the classical algebras,
although here we shall actually confine our attention to the
minimal HS gauge theory, where the matrices are one-dimensional.
We would like to emphasize, however, that \emph{the matrix
extensions have the flavor of the Chan-Paton generalizations of
open strings \cite{chanpaton}, and their consistency actually
rests of the same key features of the matrix algebras, which
suggests a natural link of the Vasiliev equations to open, rather
than to closed strings, in the tensionless limit}. More support
for this view will be presented in \cite{ps}.

The Vasiliev construction \cite{vd} requires, as in \cite{v1}, a
pair of master fields, a $1$-form $\widehat A$ and a $0$-form
$\widehat \Phi$. It rests on a set of integrable constraints on
the exterior derivatives $d\widehat A$ and $d\widehat\Phi$ that,
when combined with Lorentz-trace conditions on the component
fields, give the equations of motion within the framework of
unfolded dynamics \cite{unfolded}, whereby a given field is
described via an infinity of distinct, albeit related, tensor
fields. Notice that here \emph{all} curvatures are constrained, so
that the condition on the 2-form curvature embodies generalized
torsion constraints and identifies generalized Riemann curvatures
with certain components of $\widehat \Phi$, while the constraint
on the 1-form curvature subsumes the Bianchi identities as well as
the very definition of an unfolded scalar field. The presence of
trace parts allows in this vector construction both an
``off-shell'' formulation, that defines the curvatures, and an
``on-shell'' formulation, where suitable trace conditions turn
these definitions into dynamical equations. This is in contrast
with the original 4D spinor construction \cite{v1}, that is only
in ``on-shell'' form, since the special properties of $SL(2,C)$
multi-spinors make its 0-form components inevitably traceless.

Aside from leading to an expansion of the 0-form master field in
terms of traceful, and thus off-shell, Riemann tensors, the vector
oscillators also introduce an $Sp(2,R)$ redundancy. The crucial
issue is therefore how to incorporate the trace conditions for
on-shell Riemann tensors and scalar-field derivatives into
suitable $Sp(2,R)$-invariance conditions on the master fields.
Here, and in a more extensive paper that we hope to complete soon
\cite{s3}, we would like to propose that the Vasiliev equations of
\cite{vd} be supplemented with the \emph{strong}
$Sp(2,R)$-invariance condition
\be {\widehat{K}}_{ij} \ \star \ {\widehat \Phi} \ = \ 0 \ ,
\label{strong}\ee
where the ${\widehat{K}}_{ij}$ are fully interacting $Sp(2,R)$
generators, while subjecting $\widehat A$ only to the \emph{weak}
condition
\be \widehat D \widehat K_{ij} \ = \ 0 \ . \label{strongA} \ee
Whereas one could consider alternative forms of the strong
projection condition \eq{strong}, involving various higher-order
constructs of the $Sp(2,R)$ generators, the above linear form is
motivated to some extent by the Lagrangian formulation of
tensionless strings, or rather, string bits \cite{ps}, as well as
by related previous works of Bars and others on two-time physics
and the use of Moyal products in String Field Theory
\cite{Bars:2001um,Bars:2001ma}. There is one crucial difference,
however, in that the Vasiliev equations (\ref{strong}) involve an
additional set of vector oscillators, called $Z^{iA}$ in the
following, whose origin in the tensionless limit will be made more
transparent in \cite{ps}. Further support for \eq{strong} is
provided by previous constructions of linearized 5D and 7D HS
gauge theories based on commuting spinor oscillators \cite{5d,7d},
where the linearized trace conditions are naturally incorporated
into strong $U(1)$ and $SU(2)$ projections of the corresponding
Weyl 0-forms. As we shall see, the projection (\ref{strong})
imposes trace conditions leading to correct mass-terms in the
linearized field equations for the Weyl tensors and the scalar
field, while the resulting gauge field equations exhibit a mixing
phenomenon. To wit, the Einstein metric arises in an admixture
with the scalar field, that can be resolved by a Weyl rescaling.

The strong $Sp(2,R)$ projection \eq{strong} introduces
non-polynomial redefinitions of the linearized zero-form $\Phi$
via ``dressing functions'', similar to those that first appeared
in the spinorial constructions of \cite{5d,7d}, corresponding to a
projector $M$, via $\Phi=M\star C$ as discussed in Section 4.3. It
turns out that this projector is singular in the sense that
$M\star M$ is divergent, and thus $M$ is not normalizable
\cite{mishacubic,misha}. This poses a potential obstruction to a
well-defined curvature expansion of the Vasiliev equations based
on the vector oscillators. We have started to address this issue,
and drawing on the similar, if simpler, $U(1)$ projection of
\cite{5d,mishacubic}, we discuss how, in principle, one could
extract nonetheless a finite curvature expansion. At the present
time, however, it cannot be fully excluded that the
vector-oscillator formulation of \cite{vd} contains a pathology,
so that consistent interactions would only exist in certain low
dimensions fixed by the HS algebra isomorphisms to
spinor-oscillator realizations outlined in Section 2, although
this is unlikely. In fact, the presence of such limitations would
be in line with the suggestion that the Vasiliev equations be
somehow related to String Field Theory, and might therefore retain
the notion of a critical dimension, but no such constraints are
visible in the tensionless limit of free String Field Theory, that
rests on a contracted form of the Virasoro algebra where the
central charge has disappeared altogether. A final word on the
matter would require a more thorough investigation of the actual
interactions present in the $Sp(2,R)$ system, to which we plan to
return soon \cite{s3}.

An additional result discussed here concerns the existence of a
direct link between the linearized equations and the geometric
formulation with \emph{traceful} gauge fields and parameters of
\cite{Francia:2002aa,Sagnotti:2003qa}. The strong condition
\eq{strong} on $\widehat \Phi$, together with the weak condition
\eq{strongA} on $\widehat A$, implies indeed that the spin-$s$
gauge field equations embodied in the Vasiliev equations take the
local compensator form
\be \cF_{a_1...a_s}=\nabla_{(a_1}\nabla_{a_2}\nabla_{a_3}
\a_{a_4...a_s)} +\mbox{AdS covariantizations} \label{eq:fr1}\ ,\ee
where $\cF$ denotes the Fronsdal operator \cite{fronsdal} and
$\alpha$ is a spin-$(s-3)$ compensator, that here often differs
from its definition in \cite{Francia:2002aa,Sagnotti:2003qa} by an
overall normalization. This result reflects the link, first
discussed by Bekaert and Boulanger \cite{bb}, between the
Freedman-de Wit connections \cite{fdw} and the compensator
equations of \cite{Francia:2002aa,Sagnotti:2003qa}. These
non-Lagrangian compensator equations are equivalent, in their
turn, to the non-local geometric equations of
\cite{Francia:2002aa}, where the Fronsdal operator is replaced by
a higher-spin curvature.

By and large, we believe that a thorough understanding of HS gauge
theories can be instrumental in approaching String Theory at a
deeper conceptual level, since the HS symmetry already implies a
far-reaching extension of the familiar notion of spacetime. In
particular, the unfolded formulation \cite{unfolded} embodies an
unusually large extension of diffeomorphism invariance, in that it
is intrinsically independent of the notion of spacetime
coordinates, and therefore lends itself to provide a natural
geometric basis for translation-like gauge symmetries, somehow in
the spirit of the discussion of supertranslations in \cite{susy}.

The plan of this article is as follows. We begin in Section 2 by
relating the HS symmetries to tensionless limits of strings and
branes in AdS, and continue in Section 3 with the off-shell
definitions of the curvatures, to move on, in Section 4, to the
on-shell theory and the linearized field equations, including
their compensator form. Finally, in Section 5 we examine the
singular projector and other related non-polynomial objects,
primarily for the simpler 5D $U(1)$ case, and suggest how one
could arrive at a finite curvature expansion.

This article covers part of the lectures delivered by one of us
(P.S.) at the First Solvay Workshop on Higher-Spin Gauge Theories.
A group of students was expected to edit the complete lectures and
to co-author the resulting manuscript with the speaker, as was the
case for other Solvay talks, but this arrangement turned out not
to be possible in this case. We thus decided to write together
this contribution, combining it with remarks made by the other
authors at other Meetings and with some more recent findings, in
order to make some of the results that were presented in Brussels
available. Hopefully, we shall soon discuss extensively these
results in a more complete paper \cite{s3}, but a fully
satisfactory analysis cannot forego the need for a better grasp of
the non-linear interactions in the actual $SP(2,R)$ setting. The
remaining part of the lectures of P.S. was devoted to the relation
between bosonic string bits in the low-tension limit, minimal HS
algebras and master equations. Their content will be briefly
mentioned in the next section, but will be published elsewhere
\cite{ps}, together with the more recent results referred to
above, that were obtained in collaboration with J. Engquist.

%%%%%%%%%%%%%%%%%%%%%%%%%%%%%%%%%%%%%%%%%%%%%%%%%%%%%%%%%%%%%%%

\section{HIGHER SPINS AND TENSIONLESS LIMITS}\label{sec:t0}

%%%%%%%%%%%%%%%%%%%%%%%%%%%%%%%%%%%%%%%%%%%%%%%%%%%%%%%%%%%%%%%

HS symmetries naturally arise in the tensionless limits of strings
and branes in AdS, and there is a growing literature on this very
interesting subject, for instance
\cite{oldlindstrom,pashnev,thorn,Bars:2001um,Sundborg:2000wp,Bars:2001ma,Sezgin:2002rt,
deser,Sagnotti:2003qa,petkou,Lindstrom:2003mg,Bianchi:2003wx,Bakas:2004jq}.
The weakly coupled description of a $p$-brane with small tension
is a system consisting of discrete degrees of freedom, that can be
referred to as ``bits'', each carrying finite energy and momentum
\cite{thorn}. In the simplest case of a bit propagating in
$AdS_D$, one finds the locally $Sp(2,R)$-invariant Lagrangian
\be S\ =\ \int ~ DY^{iA} ~Y_{iA}\ ,\label{SN}\ee
where $Y^{iA}$ $(i=1,2)$ are coordinates and momenta in the
$(D+1)$-dimensional embedding space with signature
$\eta_{AB}=(--+...+)$ and $DY^{i A}=dY^{i A} +\Lambda^{ij}Y_j^A$
denotes the $Sp(2,R)$-covariant derivative along the world line.
The local $Sp(2,R)$ symmetry, with generators
\be K_{ij}\ = \ \frac12 \, Y^A_{i} \, Y_{jA} \label{SP2n} \ ,\ee
embodies the $p$-brane $\tau$-diffeomorphisms and the constraints
associated with the embedding of the conical limit of $AdS_D$ in
the target space. Multi-bit states with fixed numbers of bits are
postulated to be exact states in the theory, in analogy with the
matrix-model interpretation of the discretized membrane in flat
space. A more detailed discussion of the open-string-like quantum
mechanics of such systems will be presented in \cite{ps}.

A crucial result is that the state space of a single bit,
\be {\cal H}_{\mbox{$1$-bit}}\ =\ \{~\ket{\Psi}~:\ K_{ij}\star
\ket{\Psi}=0~\}\ ,\ee
where $\star$ denotes the non-commutative and associative
oscillator product, coincides with the CPT self-conjugate scalar
\emph{singleton} \cite{Bars:2001ma,misha,ps}
\be {\cal H}_{\mbox{$1$-bit}}=D(\epsilon_0;\{0\})\oplus
\widetilde{D}(\epsilon_0;\{0\})\ ,\qquad \epsilon_0=\frac{1}{2} \,
(D-3) \ ,\ee
where $D(E_0;S_0)$ and $\widetilde{D}(E_0;S_0)$ denote
\footnote{We denote $SL(N)$ highest weights by the number of boxes
in the rows of the corresponding Young tableaux,
$(m_1,\dots,m_n)\equiv (m_1,\dots ,m_n,0,\dots,0)$, and
$SO(D-1,T)$ highest weights by
$\{m_1,\dots,m_n\}\equiv\{m_1,\dots,m_n,0,\dots,0\}$, for $N=D+1$
or $D$, and $T=1$ or $2$. The values of $N$ and $T$ will often be
left implicit, but at times, for clarity, we shall indicate them
by a subscript.} lowest and highest weight spaces of the
$SO(D-1,2)$ generated by
\be M_{AB}\ =\ \frac12 \, Y^i_A \, Y_{iB}\ . \ee

The naively defined norm of the 1-bit states diverges, since the
lowest/highest weight states $\ket{\Omega_\pm}$ are ``squeezed''
$Y^i_A$-oscillator excitations \cite{misha,ps}. As we shall see, a
related issue arises in the corresponding HS gauge theory, when it
is subject to the strong $Sp(2,R)$-invariance condition
\eq{strong}.

The single-bit Hilbert space is irreducible under the combined
action of $SO(D-1,2)$ and the discrete involution $\pi$ defined by
$\pi(\ket{\Omega_\pm})=\ket{\Omega_\mp}$, and by
\be \pi(P_a)\ =\ -P_a\ ,\qquad \pi(M_{ab})\ =\ M_{ab}\
,\label{pi1}\ee
where $P_a=V_a{}^AV^B M_{AB}$ and $M_{ab}=V_a{}^A V_b{}^BM_{AB}$
are the (A)dS translation and rotation generators, and the
tangent-space Lorentz index $a$ is defined by $X_a=V_a{}^AX_A$ and
$X=V^A X_A$, where $X_A$ is any $SO(D-1,2)$ vector and
$(V_a{}^A,V^A)$ is a quasi-orthogonal embedding matrix subject to
the conditions
\be V_a{}^A \; V_b{}^B\; \eta_{AB}\, =\, \eta_{ab}\ ,\qquad
V_a{}^A \; V_A\, =\, 0\ ,\qquad V^A \; V_A\, =\, -\; 1\ .
\label{embedding} \ee

The AdS generators, however, do not act transitively on the
singleton weight spaces. The smallest Lie algebra with this
property is the \emph{minimal HS extension $ho_0(D-1,2)$ of
$SO(D-1,2)$} defined by \footnote{This algebra is denoted by
$hu(1/sp(2)[2,D-1])$ in
\cite{vd}.}\cite{Gunaydin:1989um,eastwood,vd,misha,ps,s3}
\be ho_0(D-1,2)\ =\ Env_1(SO(D-1,2))/{\cal I}\ , \label{ho} \ee
where $Env_1(SO(D-1,2))$ is the subalgebra of $Env(SO(D-1,2))$
elements that are odd under the $\tau$-map
\be \tau(M_{AB})\ =\ -M_{AB}\ ,\label{tau1}\ee
that may be regarded as an analog of the transposition of matrices
and acts as an anti-involution on the enveloping algebra,
\be \tau(P\star Q)\ =\ \tau(Q)\star \tau(P)\ ,\qquad P,Q\in
Env(SO(D-1,2))\ , \ee
and ${\cal I}$ is the subalgebra of $Env_1(SO(D-1,2))$ given by
the annihilating ideal of the singleton
\be {\cal I}\ =\ \{P\in Env_1(SO(D-1,2))~:~ P\star
\ket{\Psi}=0\mbox{ for all $\ket{\Psi}\in
D(\epsilon_0;\{0\})$}~\}\ .\ee

The minimal HS algebra has the following key features
\cite{misha,ps}:
\begin{enumerate}
\item[1.] It admits a decomposition into \emph{levels} labelled
by irreducible finite-dimensional $SO(D-1,2)$ representations
$\{2\ell+1,2\ell+1\}$ for $\ell=0,1,2,...$, with the $0$-th level
identified as the $SO(D-1,2)$ subalgebra.
\item[2.] It acts transitively on the scalar singleton weight
spaces.
\item[3.] It is a minimal extension of $SO(D-1,2)$, in
the sense that if $SO(D-1,2)\subseteq \cL\subseteq ho_0(D-1,2)$
and $\cL$ is a Lie algebra, then either $\cL=SO(D-1,2)$ or
$\cL=ho_0(D-1,2)$.
\end{enumerate}

One can show that these features determine uniquely the algebra,
independently of the specific choice of realization \cite{ps}. In
particular, in $D=4,5,7$ this implies the isomorphisms
\cite{misha,ps}
\be ho_0(3,2)\simeq hs(4)\ ,\qquad ho_0(4,2)\simeq hs(2,2)\
,\qquad ho_0(6,2)\simeq hs(8^*)\ ,\label{iso}\ee
where the right-hand sides denote the spinor-oscillator
realizations.

From the space-time point of view, the dynamics of tensionless
extended objects involves processes where multi-bit states
interact by creation and annihilation of pairs of bits. Roughly
speaking, unlike ordinary multi-particle states, multi-bit states
have an extended nature that should reflect itself in a
prescription for assigning weights $\hbar^{L}$ to the amplitudes
in such a way that ``hard'' processes involving many simultaneous
collisions be suppressed with respect to ``soft'' processes. This
prescription leads to a (1+1)-dimensional topological $Sp$-gauged
$\sigma$-model {\it \`a la} Cattaneo-Felder \cite{CF}, whose
associated Batalin-Vilkovisky master equation have a structure
that might be related to the Vasiliev equations \cite{ps}.

The $2$-bit states are of particular interest, since it is natural
to expect, in many ways, that their classical self-interactions
form a \emph{consistent truncation} of the classical limit of the
full theory. The 2-bit Hilbert space consists of the symmetric (S)
and anti-symmetric (A) products of two singletons, that decompose
into \emph{massless one-particle states} in AdS with even and odd
spin \cite{Flato:1978qz,misha,ps}
\be D(\epsilon_0;\{0\})\otimes D(\epsilon_0;\{0\})\ =\
\left[\bigoplus_{\tiny s=0,2,\dots}\!\!D(s+2
\epsilon_0;\{s\})\right]_{\rm S} \oplus \left[\bigoplus_{\tiny
s=1,3,\dots}\!\! D(s+2 \epsilon_0;\{s\})\right]_{\rm A}\ . \ee
The symmetric part actually coincides with the spectrum of the
\emph{minimal bosonic HS gauge theory} in $D$ dimensions, to which
we now turn our attention.

%%%%%%%%%%%%%%%%%%%%%%%%%%%%%%

\section{THE OFF-SHELL THEORY}

%%%%%%%%%%%%%%%%%%%%%%%%%%%%%%

\subsection{General Set-Up}

%%%%%%%%%%%%%%%%%%%%%%%%%%%%

The HS gauge theory based on the minimal bosonic HS algebra
$ho_0(D-1,2)$, given in \eq{ho}, is defined by a set of
constraints on the curvatures of an \emph{adjoint one form}
$\widehat A$ and a \emph{twisted-adjoint zero form}
$\widehat\Phi$. The off-shell master fields are defined by
expansions in the $Y^i_A$ oscillators, obeying a \emph{weak}
$Sp(2,R)$ invariance condition that will be defined in
eq.~\eq{sp2rinv2} below. The constraints reduce drastically the
number of independent component fields without implying any
on-shell field equations. The independent fields are a real scalar
field, arising in the master 0-form, a metric vielbein and an
infinite tower of symmetric rank-$s$ tensors for $s=4,6,\dots$,
arising in the master $1$-form. The remaining components are
auxiliary fields: in $\widehat A_\mu$ one finds the Lorentz
connection and its HS counterparts, that at the linearized level
reduce to the Freedman-de Wit connections in a suitable gauge to
be discussed below; in $\widehat \Phi$ one finds the derivatives
of the scalar field, the spin $2$ Riemann tensor, its higher-spin
generalizations, and all their derivatives. Therefore, the minimal
theory is a HS generalization of Einstein gravity, with infinitely
many bosonic fields but no fermions, and without an internal gauge
group. We have already stressed that the minimal model admits
generalizations with internal symmetry groups, that enter in way
highly reminiscent of Chan-Paton groups \cite{chanpaton} for open
strings, but here we shall confine our attention to the minimal
case.

The curvature constraints define a Cartan integrable system, a
very interesting construction first introduced in supergravity, in
its simplest non-conventional setting with 1-forms and 3-forms, by
D'Auria and Fr\'e \cite{dfre}. Any such system is gauge invariant
by virtue of its integrability, and is also manifestly
diffeomorphism invariant since it is formulated entirely in terms
of differential forms. The introduction of twisted-adjoint zero
forms, however, was a key contribution of Vasiliev
\cite{unfolded}, that resulted in the emergence of the present-day
\emph{unfolded formulation}. The full HS gauge theory \emph{a
priori} does not refer to any particular space-time manifold, but
the introduction of a $D$-dimensional bosonic spacetime ${\cal
M}_D$ yields a weak-field expansion in terms of recognizable
tensor equations. An illustration of the general nature of this
setting is provided in \cite{susy}, where 4D superspace
formulations are constructed directly in unfolded form picking a
superspace as the base manifold.

A key technical point of Vasiliev's construction is an internal
noncommutative $Z$-space $\cM_Z$, that, from the space-time
viewpoint, may be regarded as a tool for obtaining a highly
non-linear integrable system with 0-forms on a commutative
space-time $\cM_D$ \cite{v1} from a simple integrable system on a
non-commutative extended space. Although apparently {\it ad hoc},
this procedure has a rather precise meaning within the BRST
formulation of the phase-space covariant treatment of bits {\it
\`a la} Kontsevich-Cattaneo-Felder \cite{CF}, as will be discussed
in \cite{ps}.

%%%%%%%%%%%%%%%%%%%%%%%%%%%%%%%%%%%%%%%%%%%%%%%%%%%%%%%%%%%%%%%

\subsection{Oscillators and Master Fields}
\label{sec:osc}

%%%%%%%%%%%%%%%%%%%%%%%%%%%%%%%%%%%%%%%%%%%%%%%%%%%%%%%%%%%%%%%

Following Vasiliev \cite{vd}, we work with bosonic oscillators
$Y^i_A$ and $Z^i_A$, where $A=-1,0,1,\dots,D-1$ labels an
$SO(D-1,2)$ vector and $i=1,2$ labels an $Sp(2,R)$ doublet. The
oscillators obey the associative $\star$-product algebra
\bea Y_i^A\star Y_j^B  &=& Y_i^A Y_j^B+i\e_{ij}\eta^{AB}\ ,\quad
Y_i^A\star Z_j^B\ =\ Y_i^A Z_j^B - i\e_{ij}\eta^{AB}\
,\label{yystar}
\\[10pt]
Z_i^A\star Y_j^B  &=& Z_i^A Y_j^B + i\e_{ij}\eta^{AB}\ ,\quad
Z_i^A\star Z_j^B\ =\ Z_i^A Z_j^B -i\e_{ij}\eta^{AB}\label{zzstar}
\ , \eea
where the products on the right-hand sides are Weyl ordered,
\emph{i.e.} symmetrized, and obey the conditions
\be (Y^i_A)^\dagger \ =\ Y^i_A \qquad {\rm and} \qquad
(Z^i_A)^\dagger\ =\ -Z^i_A \ . \ee
It follows that the $\star$-product of two Weyl-ordered
polynomials $\widehat f$ and $\widehat g$ can be defined by the
integral
\be \widehat f (Y,Z)\star \widehat g(Y,Z)\ =\ \int {d^{2(D+1)}S\,
d^{2(D+1)}T\over (2\pi)^{2(D+1)}} \, \widehat f (Y+S,Z+S) \,
\widehat g (Y+T,Z-T) \, e^{iT^{iA}S_{iA}}\ , \label{star} \ee
where $S$ and $T$ are real unbounded integration variables.

The master fields of the minimal model are a $1$-form and a
$0$-form
\be \widehat A\ = \ dx^\m \widehat A_\m(x,Y,Z)+dZ^{iA} \widehat
A_{iA}(x,Y,Z)\ ,\qquad \widehat \Phi\ =\ \widehat\Phi(x,Y,Z)\ ,
\la{ic1}\ee
subject to the conditions defining the adjoint and twisted-adjoint
representations,
\be\t(\widehat A)\ =\ -\widehat A\ ,\qquad\widehat A^\dagger\ =\
-\widehat A\ ,\quad\quad\t(\widehat \Phi)\ =\ \pi(\widehat \Phi)\
,\qquad \widehat \Phi^\dagger\ =\ \pi(\widehat \Phi)\
,\label{phihat}\ee
where $\pi$ and $\tau$, defined in \eq{pi1} and \eq{tau1}, act on
the oscillators as
\bea \pi(\widehat f(x^\m,Y^i_a,Y^i,Z^i_A,Z^i))&=& \widehat f(x^\m,
Y^i_a,-Y^i,Z^i_a,-Z^i)\ ,\label{pi}\\[5pt] \t(\widehat
f(x^\m,Y^i_A,Z^i_A))&=&\widehat f(x^\m,iY^i_A,-iZ^i_A)\ .
\label{tau}\eea
The $\pi$-map can be generated by the $\star$-product with the
hermitian and $\tau$-invariant oscillator construct
\be \kappa\ =\ e^{i Z^i Y_i}\ =\ (\kappa)^\dagger\ =\
\tau(\kappa)\ ,\label{kappa}\ee
such that
\be \kappa\star \widehat f(Y^i,Z^i)~=~\pi(\widehat
f(Y^i,Z^i))\star\kappa~=~\kappa\widehat f(Z^i,Y^i)\
.\label{kappaf}\ee

Strictly speaking, $\kappa$ lies outside the domain of ``arbitrary
polynomials'', for which the integral representation \eq{star} of
the $\star$-product is obviously well-defined. There is, however,
no ambiguity in \eq{kappaf}, in the sense that expanding $\kappa$
in a power series and applying \eq{star} term-wise, or making use
of the standard representation of the Dirac $\delta$ function, one
is led to the same result. This implies, in particular, that
$\kappa\star\kappa=1$, so that $\kappa\star\widehat
f\star\kappa=\pi(\widehat f)$, although the form \eq{kappaf} will
be most useful in the formal treatment of the master constraints.
It is also worth stressing that the $n$-th order curvature
corrections, to be discussed later, contain $n$ insertions of
exponentials, of the type \cite{v1}
\be \cdots\star\kappa(t_1)\star \ldots \star \kappa(t_n)\cdots \
,\qquad \kappa(t)~=~e^{itZ^iY_i}\ ,\quad t_i\in[0,1]\
,\label{kappat}\ee
that are thus well-defined\footnote{We are grateful to M. Vasiliev
for an extensive discussion on these points during the Brussels
Workshop.}, and can be expanded term-wise, as was done for
instance to investigate the second-order scalar corrections to the
stress energy tensor in \cite{Kristiansson:2003xx}, or the scalar
self-couplings in \cite{Sezgin:2003pt}.

%%%%%%%%%%%%%%%%%%%%%%%%%%%%%%%%%%%%%%%%%%%%%%%%%%%%%%%%%%%%%%%

\subsection{Master Constraints}

%%%%%%%%%%%%%%%%%%%%%%%%%%%%%%%%%%%%%%%%%%%%%%%%%%%%%%%%%%%%%%%

The off-shell minimal bosonic HS gauge theory is defined by

\begin{itemize}

\item[i)] the integrable curvature constraints
\be \widehat F\ =\ {i\over 2} \, dZ^i\wedge dZ_i \, \widehat
\Phi\star \kappa\ ,\quad\quad \widehat D \widehat \Phi\ =\ 0\ ,
\label{master} \ee
where $Z^i=V^A Z_A^i$, the curvature and the covariant derivative
are given by
\be \widehat F\ =\ d\widehat A+\widehat A\star\wedge \widehat A\
,\quad\quad \widehat D \widehat \Phi\ =\ d\widehat \Phi +[\widehat
A, \widehat \Phi]_\pi\ , \label{m2} \ee
and the $\pi$-twisted commutator is defined as
\be [\widehat f,\widehat g]_\pi=\widehat f\star \widehat
g-\widehat g\star \pi(\widehat f) \ . \ee
The integrability of the constraints implies their invariance
under the general gauge transformations
\be \d_{\widehat\e} \widehat A\ =\ \widehat D\widehat \e\
\label{da}\ , \quad\quad \d_{\widehat\e} \widehat \Phi\, = \,
-[\widehat \e,\widehat\Phi]_\pi\ , \label{df} \ee
where the covariant derivative of an adjoint element is defined by
$ D\widehat \e=d\widehat \e+\widehat A\star\widehat \e-\widehat
\e\star \widehat A$;

\item[ii)] the invariance of the master fields under global
$Sp(2,R)$ gauge transformations with $\widehat\e(\l) = {i\over
2}\l^{ij}\widehat K_{ij}$, where the $\l^{ij}$ are constant
parameters, and \cite{vd}
\be \widehat K_{ij}\ =\ K_{ij}+{1\over 2}\left( \widehat S^A_{(i}
\star \widehat S_{j)A}-Z^A_iZ_{jA} \right)\; ,\ \  \widehat S^A_i
\equiv Z^A_i-2i\widehat A^A_i\ , \label{kijhat} \ee
where $K_{ij}$ are the $Sp(2,R)$ generators of the linearized
theory, defined in \eq{SP2n}.
\end{itemize}

The $Sp(2,R)$ invariance conditions can equivalently be written in
the form
\be [\widehat K_{ij},\widehat \F]_\pi\ =\ 0\ ,\qquad \widehat
D\widehat K_{ij}\ =\ 0\ . \label{sp2rinv2} \ee
These conditions remove all component fields that are not singlets
under $Sp(2,R)$ transformations. We also stress that the $\widehat
S\star\widehat S$ terms play a crucial role in the $Sp(2,R)$
generators $\widehat K_{ij}$ of \eq{kijhat}: without them all
$Sp(2,R)$ indices originating from the oscillator expansion would
transform canonically but, as shown in \cite{vd}, they guarantee
that the same holds for the doublet index of $\widehat A_{A\,i}$.

%%%%%%%%%%%%%%%%%%%%%%%%%%%%%%%%%%%%%%%%%%%%%%%%%%%%%%%%%%%%%%%%%%%%%%%%%%%%%%%%%%%%%

\subsection{Off-Shell Adjoint and Twisted-Adjoint Representations\label{sec:2.3}}

%%%%%%%%%%%%%%%%%%%%%%%%%%%%%%%%%%%%%%%%%%%%%%%%%%%%%%%%%%%%%%%%%%%%%%%%%%%%%%%%%%%%%

The gauge transformations \eq{df} are based on a rigid Lie algebra
that we shall denote by $ho(D-1,2)$, and that can be defined
considering $x$- and $Z$-independent gauge parameters.
Consequently, this algebra is defined by \cite{vd}
\be ho(D-1,2)\ =\ \left\{Q (Y):\ \t(Q )=Q ^\dagger=-Q \ ,
\quad [K_{ij},Q]_\star=0\right\}\ ,\label{hsoff} \ee
with Lie bracket $[Q,Q']_\star$, and where the linearized
$Sp(2,R)$ generators are defined in \eq{SP2n}. The $\t$-condition
implies that an element $Q $ admits the level decomposition $Q
=\sum_{\ell=0}^\infty Q _\ell$, where $Q _\ell(\l Y)=\l^{4\ell+2}Q
_\ell(Y)$ and $[K_{ij},Q _\ell]_\star=0$. As shown in \cite{vd},
the $Sp(2,R)$-invariance condition implies that $Q_{\ell}(Y)$ has
a $Y$-expansion in terms of (traceless) $SO(D-1,2)$ tensors
combining into single $SL(D+1)$ tensors with highest weights
corresponding to Young tableaux of type $(2\ell+1,2\ell+1)$:
\bea Q_\ell &=& {1\over 2i}\, Q_{A_1\dots A_{2\ell+1},B_1\dots
B_{2\ell+1}}^{ (2\ell+1,2\ell+1)}M^{A_1B_1}\cdots
M^{A_{2\ell+1}B_{2\ell+1}}\label{plev1}\w2
&=&{1\over 2i}\sum_{m=0}^{2\ell+1}Q_{a_1\dots a_{2\ell+1}b_1\cdots
b_{m}}^{(2\ell+1,m)}M^{a_1b_1}\cdots M^{a_mb_m}P^{a_{m+1}}\cdots
P^{a_{2\ell+1}}\ , \nn\eea
where the products are Weyl ordered and the tensors in the second
expression are labelled by highest weights of $SL(D)$. To
reiterate, the algebra $ho(D-1,2)$ is a reducible HS extension of
$SO(D-1,2)$, whose $\ell$-th level generators fill a {\it
finite-dimensional} reducible representation of $SO(D-1,2)$. In
the next section we shall discuss how to truncate the trace parts
at each level in $ho(D-1,2)$ to define a subalgebra of direct
relevance for the on-shell theory.

The 0-form master field $\Phi$, defined in \eq{ic} below, belongs
to the twisted-adjoint representation $T[ho(D-1,2)]$ of
$ho(D-1,2)$ associated with the rigid covariantization terms in
$\widehat D_\m\widehat\F|_{Z=0}$. Consequently,
\be T[ho(D-1,2)]~=~\left\{ R (Y):\ \t(R )=R ^\dagger=\pi(R )\
,\quad [K_{ij},R ]_\star=0\right\}\ ,\ee
on which $ho(D-1,2)$ acts via the $\pi$-twisted commutator $\d_\e
R=-[\e,R]_\pi$. A twisted-adjoint element admits the level
decomposition $R = \sum_{\ell=-1}^{\infty} R _{\ell}$, where $R
_\ell$ has an expansion in terms of $SL(D)$ irreps as
($\ell\geq-1$)
\be R _{\ell} \ = \ \sum_{k=0}^\infty {1\over k!}\ R
^{(2\ell+2+k,2\ell+2)}_{a_1\dots a_{2\ell+2+k},b_1\dots
b_{2\ell+2}} \ M^{a_1b_1}\cdots
M^{a_{2\ell+2}b_{2\ell+2}}P^{a_{2\ell+3}}\cdots
P^{a_{2\ell+2+k}}\; . \label{rlev}\ee
It can be shown that \emph{each level forms a separate, infinite
dimensional reducible $SO(D-1,2)$ representation}, that includes
an infinity of trace parts that will be eliminated in the on-shell
formulation. In particular, $R _{-1}$ has an expansion in terms of
$SL(D)$ tensors carrying the same highest weights as an off-shell
scalar field and its derivatives, while each of the $R _{\ell}$,
$\ell\geq 0$, consists of $SL(D)$ tensors corresponding to an
off-shell spin-$(2\ell+2)$ Riemann tensor and its derivatives.

%%%%%%%%%%%%%%%%%%%%%%%%%%%%%%%%%%%%%%%%%%%%%%%%%%%%%%%%%%%%%%%%%%%%%%

\subsection{Weak-Field Expansion and Linearized Field Equations \label{sec:2.4}}

%%%%%%%%%%%%%%%%%%%%%%%%%%%%%%%%%%%%%%%%%%%%%%%%%%%%%%%%%%%%%%%%%%%%%

The master constraints \eq{master} can be analyzed by a weak-field
expansion, which will sometimes be referred to as a {\it
perturbative expansion}, in which the scalar field, the HS gauge
fields and all curvatures (including the spin-two curvature) are
indeed treated as weak. One can then start from the initial
conditions
\be A_\m\ =\ \widehat A_\m|_{Z=0}\ ,\qquad \F\ =\ \widehat
\Phi|_{Z=0}\ ,\label{ic}\ee
fix a suitable gauge and solve for the $Z$-dependence of $\widehat
A$ and $\widehat \Phi$ order by order in $\Phi$ integrating the
constraints
\be {\widehat D}_i\widehat \Phi\ =\ 0\ , \qquad {\widehat F}_{ij}\
=\ -i\e_{ij} {\widehat \Phi}\star\k\ ,\qquad {\widehat F}_{i\m}\
=\ 0\ , \label{master1}\ee
thus obtaining, schematically,
\be {\widehat\Phi}\ =\ {\widehat\Phi}(\Phi)\ ,\qquad {\widehat
A}_\mu\ =\ {\widehat A}_\mu(\Phi,A_\mu)\ ,\qquad {\widehat A}_i\
=\ {\widehat A}_i (\Phi)\ . \ee
Substituting these solutions in the remaining constraints
evaluated at $Z=0$ gives
\be \widehat F_{\m\n}|_{Z=0}\ =\ 0\ ,\qquad \widehat D_\m\widehat
\Phi|_{Z=0}\ =\ 0\ ,\label{m6}\ee
that describe the full non-linear off-shell HS gauge theory in
ordinary spacetime, \emph{i.e. the full set of non-linear
constraints that define its curvatures without implying any
dynamical equations}.

A few subtleties are involved in establishing the integrability in
spacetime of \eq{m6} in perturbation theory. One begins as usual
by observing that integrability holds to lowest order in $\Phi$,
and proceeds by assuming that \emph{all} the constraints hold for
\emph{all} $x^\mu$ and $Z$ to $n$-th order in $\Phi$. One can then
show that the constraint $(\widehat D_i\widehat \Phi)^{(n+1)}=0$
is an integrable partial differential equation in $Z$ for
$\widehat\Phi^{(n+1)}$, whose solution obeys $\partial_i(\widehat
D_\mu\widehat\Phi)^{(n+1)}=0$, which in its turn implies that
$(\widehat D_\mu\widehat\Phi)^{(n+1)}=0$ for all $Z$ if $(\widehat
D_\mu\widehat \Phi)^{(n+1)}|_{Z=0}=0$. Proceeding in this fashion,
one can obtain $\widehat A_i^{(n+1)}$, and then $\widehat
A_\mu^{(n+1)}$, via integration in $Z$, to show that
$\partial_i\widehat F_{\mu\nu}^{(n+1)}=0$, so that if $\widehat
F_{\mu\nu}^{(n+1)}|_{Z=0}=0$, then $\widehat F_{\mu\nu}^{(n+1)}=0$
for all $Z$. It thus follows, by induction, that once the
constraints \eq{m6} on $\widehat F_{\mu\nu}$ and $\widehat \Phi$
are imposed at $Z=0$, they hold for all $Z$, and are manifestly
integrable on spacetime, since $Z$ can be treated as a parameter.
Hence, their restriction to $Z=0$ is also manifestly integrable in
spacetime, simply because the exterior derivative
$dx^\mu\partial_\mu$ does not affect the restriction to $Z=0$.

Having obtained \eq{m6} in a $\Phi$-expansion, one can write
\be A_\m = e_\m + \o_\m + W_\m\ ,\label{emomWm}\ee
where $e_\m={1\over 2i}\, e_\m{}^aP_a$ and $\o_\m={1\over 2i}\,
\o_\m{}^{ab}M_{ab}$, and where $W_\m$ contains the HS gauge fields
residing at levels $\ell\geq 1$. We would like to stress that
until now $\mu$ has been treated as a formal curved index with no
definite intrinsic properties. Treating $e_\mu$ and $\omega_\mu$
as strong fields and referring $\mu$ explicitly to a
$D$-dimensional bosonic spacetime clearly builds a perturbative
expansion that preserves local Lorentz invariance and
$D$-dimensional diffeomorphism invariance. It would be interesting
to investigate to what extent the higher-spin geometry could be
made more manifest going beyond this choice.

To first order in the weak fields $\Phi$ and $W_\mu$, the
constraints \eq{m6} reduce to
\bea && {\mathcal  R} +{\mathcal F} \ = \ i \; e^a\wedge
e^b\left.{\partial^2\Phi\over\partial Y^{ai}\partial
Y^b_i}\right|_{Y^i=0}
\ ,\label{lin2}\\
&& \nabla\Phi + {1\over 2i}\, e^a\{P_a,\Phi\}_\star \ =\ 0\
,\label{Phieqoffshell1}\eea
where $\cR$ is the $SO(D-1,2)$-valued curvature of $E\equiv e+\o$
defined by $\cR=dE+ [E,E]_\star$, and $\cF$ is the linearized
$SO(D-1,2)$ covariant curvature defined by $\cF=dW+\{E,W\}_\star
$, and $\nabla \Phi=d\Phi +[\omega,\Phi]_\pi$ is the Lorentz
covariant derivative of $\Phi$. Since each level of the adjoint
and twisted-adjoint master fields forms a separate representation
of $SO(D-1,2)$, the linearized constraints split into independent
sets for the individual levels:
\bea \ell=0&:&\cR\ =\ -8\; i\, e^a\wedge e^b \;
\F^{(2,2)}_{ac,bd}\, M^{cd}\
,\label{gravconstr1}\\[10pt]
\ell\geq 1&:&\cF_\ell\ =\ i\; \e^a\wedge e^b\;
{\partial^2\Phi_\ell^{(2\ell+2,2\ell+2)}\over\partial
Y^{ai}\partial
Y^b_i}\ ,\label{lin3}\\[10pt]
\ell\geq -1&:& \nabla\Phi_\ell + {1\over 2i}\, e^a\;
\{P_a,\Phi_\ell\}_\star\ =\ 0\ .\label{Phieqoffshell} \eea
Note that ${\cal F}$ is to be expanded as in \eq{plev1}, and
${\Phi_\ell}$ as in \eq{rlev}. Furthermore, in order to compute
the star anticommutator in the last equation, one must use the
$Y$-expansion of all generators involved, and double contractions
contribute to this term. Note also, for example, that even if the
star commutator occurs in ${\cal R}$, using the commutation
relation between the AdS generators is not sufficient, and one
must also recall relations such as $[M_{[ab},P_{c]}]=0$, that
follow from the oscillator realization of these generators.

The first of \eq{Phieqoffshell} contains the usual torsion
constraint, and identifies $\F^{(2,2)}_{ab,cd}$ with the
$SO(D-1,2)$-covariantized Riemann curvature of $e_\mu{}^a$. As
$\F^{(2,2)}_{ab,cd}$ is traceful, these equations describe
off-shell AdS gravity: the trace of \eq{gravconstr1} simply
determines the trace part of $\F^{(2,2)}_{ab,cd}$, rather than
giving rise to the Einstein equation. This generalizes to the
higher levels, and a detailed analysis of \eq{lin3} reveals that
\cite{s3}:
\begin{itemize}
\item[i)] the gauge parameter $\e_\ell$ contains
St\"uckelberg-type shift symmetries;
\item[ii)] $W_\ell$ contains pure gauge parts and auxiliary gauge
fields that can be eliminated using shift symmetries or
constraints on torsion-like components of $\cF_\ell$,
respectively;
\item[iii)] the remaining independent components of $W_\ell$
correspond to the fully symmetric tensors
\be \phi^{(s)}_{a_1\dots a_s}\ \equiv\ e_{(a_1}{}^\m\
W^{(s-1)}_{\m,a_2\dots a_{s})}\ ,\qquad s=2\ell+2 \
;\label{phis}\ee
\item[iv)] the system is off-shell: the remaining non-torsion-like
components of $\cF_\ell$ \emph{vanish identically}, with the only
exception of the $s$-th one, that defines the generalized
(traceful) Riemann tensor of spin $s=2\ell+2$,
\be R^{(s,s)}_{a_1\dots a_s,}{}^{b_1\dots b_s}\ \equiv\
e^\m_{(a_1}{} e^{\n(b_1 }\cF^{(s-1,s-1)}_{\m\n,a_2\dots
a_s),}{}^{b_2\dots b_s)}= 4s^2\Phi^{(s,s)}_{a_1\dots
a_s,}{}^{b_1\dots b_s}\ ,\label{genRiemann} \ee
where the identification follows from \eq{lin2} and the Riemann
tensor is built from $s$ derivatives of $\phi^{(s)}$.
\end{itemize}

Turning to the $\F_\ell$-constraint \eq{Phieqoffshell}, one can
show that its component form reads ($s=2\ell+2$, $\ell\geq -1$,
$k\geq 0$):
\be \nabla_\mu \Phi^{(s+k,s)}_{a_1\dots a_{s+k},b_1\dots b_s} \ =\
{i\over 4}\, (s+k+2)
\Phi^{(s+k+1,s)}_{\mu}\hspace{-36pt}\underbrace{{}^{\phantom{()}}_{
a_1\dots a_{s+k},b_1\dots b_s}}_{(s+k,s)}+~i{(k+1)(s+k)\over
s+k+1}~ \eta^{\phantom{(}}_{\mu(a_1}\Phi^{(s+k-1,s)}_{a_2\dots
a_{s+k}),b_{1}\dots b_{s}}\ ,\label{phieq}\ee
where we have indicated a Young projection to the tableaux with
highest weight $(s+k,s)$. Symmetrizing $\mu$ and $a_1\dots a_k$
shows that $\Phi^{(s+k+1,s)}$ ($k\geq 0$) are auxiliary fields,
expressible in terms of derivatives of $\Phi^{(s,s)}$. The
$\Phi^{(s,s)}$ components are generalized (traceful) Riemann
tensors given by \eq{genRiemann} for $\ell\geq 0$, and an
independent scalar field for $\ell=-1$,
\be \phi\ \equiv\ \Phi^{(0,0)}\ .\label{phi}\ee
The remaining components of \eq{phieq}, given by the $(s+k,s+1)$
and $(s+k,s,1)$ projections, are Bianchi identities. Hence, no
on-shell conditions are hidden in \eq{phieq}.

In particular, combining the $k=0,1$ components of \eq{phieq} for
$s=0$, one finds
\be \left(\nabla^2+{D\over 2}\right)\phi\ =\ -\, {3\over
8}~\eta^{ab}~\Phi^{(2,0)}_{ab}\ ,
 \label{offshellse}\ee
which, as stated above, determines the trace part
$\eta^{ab}\Phi^{(2)}_{ab}$ of the auxiliary field $\F^{(2)}_{ab}$
rather than putting the scalar field $\phi$ on-shell. Similarly,
the $s\geq 2$ and $k=0,1$ components of \eq{phieq} yield
\be \nabla^2 \Phi^{(s,s)}_{a_1\dots a_s,b_1\dots b_s}
~+~{(D+s)\over 2}~\Phi^{(s,s)}_{a_1\dots a_s,b_1\dots b_s}\ =\
\frac{(s+2)(s+3)}{16}~\eta^{\mu\nu}\Phi_{\mu\nu\phantom{b}}^{(s+2,s)}
\hspace{-19pt}\underbrace{{}^{\phantom{()}}_{a_1\dots a_s,b_1\dots
b_s}}_{(s,s)}\ ,\label{klein} \ee
that determine the trace parts of the auxiliary fields
$\Phi^{(s+2,s)}$, rather than leading to the usual
Klein-Gordon-like equations satisfied by on-shell Weyl tensors.

In summary, the constraints \eq{master} describe an off-shell HS
multiplet with independent field content given by a tower of real
and symmetric rank-$s$ $SL(D)$ tensors with $s=0,2,4,...$,
\be \phi~,\ e_\m{}^a~,\ \phi^{(s)}_{a_1\dots a_s} \quad
(s=4,6,\dots)\ ,\ee
where the scalar field $\phi$ is given in \eq{phi}, the vielbein
$e_\m{}^a$ is defined in \eq{emomWm}, and the real symmetric HS
tensor fields are given in \eq{phis}.

%%%%%%%%%%%%%%%%%%%%%%%%%%%%%%%%%%%%%%%%%%%%%%%%%%%%%%%%%%%%%%%%%%%%%

\section{THE ON-SHELL THEORY}\label{sec:3}

%%%%%%%%%%%%%%%%%%%%%%%%%%%%%%%%%%%%%%%%%%%%%%%%%%%%%%%%%%%%%%%%%%%%%

%%%%%%%%%%%%%%%%%%%%%%%%%%%%%%%%%%%%%%%%%%%%%%%%%%%%%%%%%%%%%%%%%%%%%

\subsection{On-Shell Projection}
\label{sec:on}

%%%%%%%%%%%%%%%%%%%%%%%%%%%%%%%%%%%%%%%%%%%%%%%%%%%%%%%%%%%%%%%%%%%%%

It should be appreciated that, if the trace parts in
\eq{offshellse} and \eq{klein} were simply dropped, \emph{the
resulting masses would not coincide with the proper values for a
conformally coupled scalar and on-shell spin-$s$ Weyl tensors in
$D$ dimensions}. Hence, the trace parts contained in $\Phi$ must
be carefully eliminated, and cannot be simply set equal to zero.
To this end, as discussed in the Introduction, motivated by the
arguments based on $Sp(2,R)$-gauged noncommutative phase spaces
arising in the context of tensionless strings and two-time
physics, we would like to propose the on-shell projection
\be \widehat K_{ij}\star \widehat \Phi \ =\ 0 \ , \label{kijproj}
\ee
where $\widehat K_{ij}$ is defined in (\ref{kijhat}), be adjoined
to the master constraints of \cite{vd}
\be  \widehat F\ = \ {i\over 2}\ dZ^i\wedge dZ_i\
\widehat\Phi\star\k\ ,\qquad \widehat D\widehat\Phi \ =\ 0
\label{m4} \ . \ee
Although one can verify that the combined constraints \eq{kijproj}
and \eq{m4} remain formally integrable at the full non-linear
level, the strong $Sp(2,R)$ projection (\ref{kijproj}) should be
treated with great care, since \emph{its perturbative solution
involves non-polynomial functions of the oscillators associated
with projectors whose products can introduce divergences in higher
orders of the perturbative expansion unless they are properly
treated} \cite{misha}. It is important to stress, however, that
these singularities draw their origin from the curvature
expansion, and in Section \ref{sec:proj} we shall describe how a
finite curvature expansion might be defined.

We can thus begin by exploring the effects of (\ref{kijproj}) on
the linearized field equations and, as we shall see, the correct
masses emerge in a fashion which is highly reminiscent of what
happens in spinor formulations. There is a further subtlety,
however. According to the analysis in the previous section, if the
internal indices of the gauge fields in $\widehat A$ were also
taken to be traceless, the linearized 2-form constraint \eq{lin3}
would reduce to the Fronsdal equations \cite{fronsdal}, that in
the index-free notation of \cite{Francia:2002aa} read
\begin{equation}
{\cal F} \ = \ 0 \ \label{fr1},
\end{equation}
where the Fronsdal operator is
\begin{equation}
{\cal F} \ \equiv \ \nabla^2 \; \phi \ - \ \nabla \, \nabla \cdot
\phi \ + \ \nabla\; \nabla \, \phi^{'} \ - \ \frac{1}{L^2}\ \{
[(3-D-s)(2-s) -s] \phi + 2 g \phi^{'} \} \ ,\label{fr2}
\end{equation}
and where ``primes'' denote traces taken using the AdS metric $g$
of radius $L$. We would like to stress that this formulation is
based on \emph{doubly traceless} gauge fields, and is invariant
under gauge transformations with \emph{traceless} parameters.
However, the constrained gauge fields of this conventional
formulation should be contrasted with those present in (\ref{m4}),
namely the metric and the HS gauge fields collected in
(\ref{phis}). These fields do contain trace parts, and enforcing
the 0-form projection as in (\ref{kijproj}), as we shall see,
actually leaves the gauge fields free to adjust themselves to
their constrained sources in the projected Weyl 0-form $\widehat
\Phi$, thus extending the conventional Fronsdal formulation based
on \eq{fr1} and \eq{fr2} to the geometric formulation of
\cite{Francia:2002aa}. This is due to the fact that, once the
0-forms on the right-hand side of \eq{lin3} are constrained to be
traceless Weyl tensors, certain trace parts of the spin-$s$ gauge
fields can be expressed in terms of gradients of traceful
rank-$(s-3)$ symmetric tensors $\a_{a_1...a_{s-3}}$. As we shall
see in detail in subsection 4.4, up to the overall normalization
of $\alpha$, that is not chosen in a consistent fashion throughout
this paper, these will enter the physical spin-$s$ field equations
embodied in the Vasiliev constraints precisely as in
\cite{Francia:2002aa,Sagnotti:2003qa}. In the index-free notation
of \cite{Francia:2002aa}, the complete linearized field equations
in an AdS background would thus read
\be \cF \ = \ 3\, \nabla \nabla \nabla \a \ - \ \frac{4}{L^2}\ g\,
\nabla \a \label{eq:fr11}\ ,\label{compens}\ee
where ${\cal F}$ denotes again the Fronsdal operator. Notice,
however, that \emph{now the gauge field $\phi$ is traceful and
$\alpha$ plays the role of a compensator for the traceful gauge
transformations},
\begin{eqnarray}
&& \delta \; \phi \ = \ \nabla \Lambda \ , \nonumber \\
&& \delta \; \alpha \ = \ \Lambda^{'} \ ,
\end{eqnarray}
with $\Lambda'$ the trace of the gauge parameter $\Lambda$. When
combined with the Bianchi identity, eq.~\eq{compens} implies that
\cite{Sagnotti:2003qa}
\begin{equation}
\phi''\ =\ 4\nabla\cdot \alpha+\nabla\alpha'\ ,
\end{equation}
so that once $\alpha$ is gauged away using $\Lambda'$,
eq.~\eq{compens} reduces to the Fronsdal form. It will be
interesting to explore the role of the additional gauge symmetry
in the interactions of the model.

To reiterate, the system \eq{kijproj} and (\ref{m4}) provides a
realization {\it \`a la} Cartan of an on-shell HS gauge theory
that embodies the non-local geometric equations of
\cite{Francia:2002aa}, with traceful gauge fields and parameters,
rather than the more conventional Fronsdal form \cite{fronsdal}.

%%%%%%%%%%%%%%%%%%%%%%%%%%%%%%%%%%%%%%%%%%%%%%%%%%%%%%%%%%%%%%%%%%%%%%%%%%

\subsection{On-Shell Adjoint and Twisted-Adjoint
Representations} \label{sec:onhsa}

%%%%%%%%%%%%%%%%%%%%%%%%%%%%%%%%%%%%%%%%%%%%%%%%%%%%%%%%%%%%%%%%%%%%%%%%%%

It is important to stress that in our proposal both the on-shell
and off-shell systems contain a gauge field in the adjoint
representation of $ho(D-1,2)$, while in the on-shell system the
0-form obeys an additional constraint, the strong $Sp(2,R)$
projection condition \eq{kijproj}. The \emph{on-shell
twisted-adjoint representation} is thus defined for $D \geq 4$
by\footnote{In $D=3$ one can show that $S$ is actually a constant
\cite{misha,s3}.}
\be T_0[ho(D-1,2)]\ =\ \{S\in T[ho(D-1,2)]~:\ K_{ij}\star S=0\ ,
\quad S\star K_{ij}=0 \}\ , \label{twadjon}\ee
where the linearized $Sp(2,R)$ generators $K_{ij}$ are defined in
(\ref{SP2n}). The part of $A_\mu$ that annihilates the
twisted-adjoint representation can thus be removed from the rigid
covariantizations in $\widehat D_\mu\widehat \Phi|_{Z=0}$. It
forms an $Sp(2,R)$-invariant ideal $I(K)\subset ho(D-1,2)$
consisting of the elements generated by left or right
$\star$-multiplication by $K_{ij}$, \emph{i.e.}
\be I(K)\ =\ \left\{K^{ij}\star f_{ij}:
\t(f_{ij})=(f_{ij})^\dagger=-f_{ij},\
[K_{ij},f^{kl}]_{\star}=4i\d_{(i}^{(k}f_{j)}^{l)}\right\}\ .
\label{IK} \ee
\emph{Factoring out} the ideal $I(K)$ from the Lie bracket
$[Q,Q']_\star$, one is led to the \emph{minimal bosonic HS
algebra}
\be  ho(D-1,2)/I(K) \ \simeq \ ho_0(D-1,2) \ \equiv\
Env_1(SO(D-1,2))/{\cal I}\ , \ee
defined in \eq{ho}. The isomorphism follows from the uniqueness of
the minimal algebra \cite{ps}, and from the fact that
$ho(D-1,2)/I(K)$ shares its key properties \cite{vd,misha,ps}.
Thus, at the linearized level, the gauging of $ho_0(D-1,2)$ gives
rise, for each even spin $s$, to the canonical frame fields
\cite{mishaframe}
\be A_{\mu,A_1\dots A_{s-1},B_1\dots B_{s-1}} ^{\{s-1,s-1\}}\ =\
\left\{A_{\mu,a_1\dots a_{s-1},b_1\dots b_k}
^{\{s-1,k\}}\right\}_{_{k=0}}^{s-1}\ ,\label{canset}\ee
required to describe massless spin-$s$ degrees of freedom in the
conventional Fronsdal form.

We would like to stress, however, that at the non-linear level the
\emph{full} on-shell constraints \eq{kijproj} and \eq{m4} make use
of the larger, reducible set of off-shell gauge fields valued in
$ho(D-1,2)=ho_0(D-1,2)\oplus I(K)$. Hence, while the linearized
compensator form \eq{compens} can be simply gauge fixed to the
conventional Fronsdal form, interesting subtleties might well
arise at the nonlinear level. It would thus be interesting to
compare the interactions defined by \eq{kijproj} and \eq{m4}, to
be extracted using the prescription of Section 5.3, with those
resulting from the formulation in \cite{vd}.

The minimal algebra allows matrix extensions that come in three
varieties, corresponding to the three infinite families of
classical Lie algebras \cite{misha}. We have already stressed that
these enter in a fashion highly reminiscent of how Chan-Paton
factors enter open strings \cite{chanpaton}, and in this respect
the presently known HS gauge theories appear more directly related
to open than to closed strings. The minimal bosonic HS gauge
theory has in fact a clear open-string analog, the $O(1)$ bosonic
model described in the review paper of Schwarz in
\cite{chanpaton}. More support for this view will be presented in
\cite{ps}.

%%%%%%%%%%%%%%%%%%%%%%%%%%%%%%%%%%%%%%%%%%%%%%%%%%%%%%%%%%%%%%%

\subsubsection{Dressing Functions}

%%%%%%%%%%%%%%%%%%%%%%%%%%%%%%%%%%%%%%%%%%%%%%%%%%%%%%%%%%%%%%

The strong $Sp(2,R)$-invariance condition on the twisted-adjoint
representation $T_0[ho(D-1,2)]$, $K_{ij} \star S = 0$, or
equivalently $K_I\star S=0$, where $I$ is a triplet index, can be
formally solved letting \cite{misha}
\be S\ =\ M\star R\ ,\label{MstarR}\ee
where $R\in T[ho(D-1,2)]/I(K)$ and $M$ is a function of $K^2=K^I
K_I$ that is \emph{analytic} at the origin and satisfies
\be K_{ij}\star M\ =\ 0\ ,\qquad \tau(M)\ =\ M^\dagger\ =\ M\
.\label{Mdef}\ee
This and the normalization $M(0)=1$ imply that
\be M(K^2)\ =\ \sum_{p=0}^\infty {(-4K^2)^p\over p!}
{\C(\frac{D}{2})\over \C(\ft12(D+2p))}  \ =
 \ \Gamma\left(\frac{D}{2}\right)  \frac{J_{\frac{D}{2}-1}( 4 \;
\sqrt{K^2})}{(2 \sqrt{K^2})^{\frac{D}{2}-1}_{\phantom{()}}}\ ,
\label{Mf0}\ee
where $J$ is a Bessel function and $\Gamma$ is the Euler $\Gamma$
function. Actually, $M$ belongs to a class of {\it dressing
functions}
\be F(N;K^2)\ =\ \sum_{p=0}^\infty {(-4K^2)^p\over p!}
{\C(\ft12(N+D))\over \C(\ft12(N+D+2p))} \ =\
\Gamma\left({\n+1}\right)  \frac{J_\n( 4 \; \sqrt{K^2})}{(2
\sqrt{K^2})^\n_{\phantom{()}}}\ , \label{dressing}\ee
related to Bessel functions of order $\n=\frac{N+D-2}{2}$ and
argument $4\sqrt{K^2}$, and in particular
\be M(K^2)\ =\ F(0;K^2)\ .\ee

The dressing functions arise in the explicit level decomposition
of the twisted-adjoint element $S$ in \eq{MstarR}. In fact, for
$D\geq 4$ one finds
\be \qquad S=\sum_{\ell=-1}^\infty S_{\ell}\ ,\qquad
S_\ell=\sum_{q=0}^{\ell+1}S_{\ell,q}\ , \label{levon2} \ee
where the expansion of $S_{\ell,q}$ is given by ($s=2\ell+2$)
\cite{s3}
\bea S_{\ell,q}&=&\sum_{k=0}^\infty { d_{s,k,q}\over k!}\, S_{
a_1\dots a_{s+k},b_{2q+1}\dots b_s}^{\{s+k,s-2q\}}\,
\eta^{\phantom{\{}}_{b_1b_2}\cdots
\eta^{\phantom{\{}}_{b_{2q-1}b_{2q}}\, F({\tiny 2(s+k)};K^2)\times\nn\\
&&\quad\quad \times M^{a_1b_1}\cdots
M^{a_{s}b_{s}}P^{a_{s+1}}\cdots P^{a_{s+k}}\ .\label{rellon} \eea
The coefficients $d_{s,k,q}$ with $q\geq 1$ are fixed by the
requirement that all Lorentz tensors arise from the decomposition
of AdS tensors, $S^{\{s+k,s\}}\in S^{\{s+k,s+k\}_{D+1}}$, while
$d_{s,k,0}$ can be set equal to one by a choice of normalization
\cite{s3}. The Lorentz tensors arising in $S_{\ell,q}$ with $q
\geq 1$ are simply combinations\footnote{As we shall see in the
next subsection, this mixing has a sizable effect on the free
field equations.} of those in $S_{\ell,0}$, that constitute the
various levels of the on-shell twisted-adjoint representation
($s=2\ell+2)$:
\be S_{\ell,0}\ :\quad  \left\{ S^{\{s+k,s\}}_{a_1\dots
a_{s+k},b_1\dots b_s}~ (k=0,1,2,\dots)\right\}\ . \label{Sell0}
\ee
The $\ell$-th level forms an irreducible multiplet within the
twisted-adjoint representation of $SO(D-1,2)$. In showing this
explicitly, the key point is that the translations, generated by
$\d_\xi=\xi^a\{P_a,\cdots\}_\star$, do not mix different levels
\cite{s3}. Thus, $S_{-1,0}$ affords an expansion in terms of
Lorentz tensors carrying the same highest weights as an on-shell
scalar field and its derivatives, while the $S_{\ell,0}$,
$\ell\geq 0$, correspond to on-shell spin-$(2\ell+2)$ Weyl tensors
and their derivatives.

%%%%%%%%%%%%%%%%%%%%%%%%%%%%%%%%%%%%%%%%%%%%%%%%%%%%%%%%%%%%%%%%%%%%%%

\subsection{Linearized Field Equations and Spectrum of the Model}

%%%%%%%%%%%%%%%%%%%%%%%%%%%%%%%%%%%%%%%%%%%%%%%%%%%%%%%%%%%%%%%%%%%%%

In order to obtain the linearized field equations, it suffices to
consider the initial conditions at $Z=0$ to lowest order in the
0-form\footnote{The strong $Sp(2,\mathbf R)$ condition induces
higher-order $C$ corrections to the initial condition on $\widehat
\Phi$ \cite{s3}.}
\be\left.\widehat A_\mu\right|_{Z=0}\ =\  A_\mu\in ho(D-1,2)\
,\quad
\left.\widehat \F\right|_{Z=0}\ =\ M\star C + {\cal O}(C^2)
\label{icF}\ ,\ee
where $M \star C \in T_0[ho(D-1,2)]$. The first $C$ correction to
$\widehat A_\mu$ is then obtained integrating in $Z$ the
constraint $\widehat F_{i\mu}=0$. Expanding also in the
higher-spin gauge fields as discussed below \eq{emomWm} and fixing
suitable gauges \cite{vd,s3}, one finds that the Vasiliev
equations reduce to \cite{s3}
\bea && \cR+\cF\ =\  i\left.{\partial^2(C\star M)\over
\partial Y^{\m i}\partial Y^\n_i}\right|_{Y^i=0}\ ,\label{on1a}\w4
&& \nabla(C\star M)+{1\over 2i}e^a\{P_a,C\star M \}_\star\ =\ 0\ .
\label{on2a} \eea
%

%%%%%%%%%%%%%%%%%%%%%%

\subsubsection{${\Phi}$ Constraint and Role of
the Dressing Functions}

%%%%%%%%%%%%%%%%%%%%%

The expansion of the master field $C \star M$ involves the
dressing functions $F(N;K^2)$, as in \eq{rellon}. These, in their turn, play a crucial role in
obtaining the appropriate field equations already at the
linearized level, since they determine the trace parts of the
auxiliary fields $\Phi^{(s+2,s)}$ in \eq{klein}. Using these
traces in the iterated form of \eq{on2a}, one can show that the
field equations for the scalar $\phi=C|_{Y=0}$ and the Weyl
tensors $C^{\{s,s\}}$ are finally
 \be
 (\nabla^2-m_0^2)\, \phi\ =\ 0\ ,\qquad (\nabla^2 -
 m_s^2)\, C^{\{s,s\}}_{a_1\dots a_s,b_1\dots b_s}=0\ ,\qquad m_s^2 \
 =\
 -\, {1\over 2}\, (s+D-3)\ ,
 \label{mass}
 \ee
and contain the proper mass terms. The details of the computations
leading to this result will be given in \cite{s3}, but the effect
on the scalar equation is simple to see and can be spelled out in
detail here. Indeed, the relevant contributions to $K_{ij} \star
\Phi$ can be traced to the action of $K_{ij}$ on $(\phi +
\frac{1}{2} \, \Phi^{(2,0)}_{ab} P^a P^b)$, that leads, after
expanding the $\star$-products, to a term proportional to $K_{ij}
\left( \phi + \frac{1}{4}\, \eta^{ab} \,
\Phi^{(2,0)}_{ab}\right)$. This disappears precisely if $\eta^{ab}
\, \Phi^{(2,0)}_{ab} \, = \, - \, 4 \; \phi$, which leads, via
eq.~\eq{offshellse}, to the correct scalar mass term.

Once the mass terms are obtained, it is straightforward to
characterize the group theoretical content of the spectrum, as we
shall do in subsection 4.3.3.

%%%%%%%%%%%%%%%%%%%%%%

\subsubsection{${\cal F}$ Constraint and Mixing
Phenomenon}

%%%%%%%%%%%%%%%%%%%%%%

Since on shell and after gauge fixing ${\cF}\in ho(D-1,2)/I(K)$,
its expansion takes the form \eq{plev1}, in which the (traceful)
$SL(D)$ representations $Q^{(2\ell+1,m)}$ must be replaced by the
(traceless) $SO(D-1,1)$ representations $Q^{\{2\ell+1,m\}}$. Using
this expansion for ${\cal F}$, and the expansions \eq{levon2} and
\eq{rellon} for $C\star M$, an analysis of the ${\cF}$ constraint
\eq{on1a} shows that the physical fields at level $\ell$ present
themselves in an admixture with lower levels. Still, the
integrability of the constraints ensures that diagonalization is
possible, and we have verified this explicitly for the spin-2
field equation. Indeed, starting from \eq{on1a} one finds
 \be
 R_{\m\n}\, +\, \ft14(D-1)g_{\m\n}\
 =\ -\, \ft{16}3\nabla_{\{\m}\nabla_{\n\}}\phi
\, +\, \ft{8(D-1)}{D} \, g_{\m\n}\, \phi\ , \label{Ric1} \ee
where $R_{\mu\nu}$ denotes the spin-2 Ricci tensor for the metric
$g_{\mu\nu}=e_\mu{}^a e_{\nu\,a}$ and $\phi$ is the physical
scalar. Using the scalar field equation \eq{mass}, one can show
that the rescaled metric
\be \tilde g_{\m\n}\ =\ e^{-2u}~g_{\m\n}\ ,\qquad {\rm with} \quad
u=\ft{16}{3(D-2)}\phi\ ,\label{Weyltransf}\ee
is the Einstein metric obeying
\be \widetilde R_{\m\n}+\ft14(D-1)\tilde g_{\m\n}\ =\ 0\ \ .
\label{Ric11}\ee
It is natural to expect that this result extend to higher orders,
and that the end result be equivalent to some generalization of
the Weyl transformation \eq{Weyltransf} at the level of master
fields.

Barring this mixing problem, the Fronsdal operators at higher
levels can be sorted out from the constraint on $\cF$ in \eq{on1a}
noting that the $SO(D-1,2)$-covariant derivatives do not mix
$SO(D-1,2)$ irreps, which allows one to consistently restrict the
internal indices at level $\ell$ to the $SO(D-1,2)$ irrep with
weight $\{2\ell+1,2\ell+1\}$. Decomposing this irrep into Lorentz
tensors and using cohomological methods \cite{misha2,s3} then
reveals that the curvatures are properly on-shell, although the
precise form of the resulting field equations, as we have
anticipated, requires a more careful elimination of the ideal
gauge fields, that will be discussed in subsection 4.4.

%%%%%%%%%%%%%%%%%%%%%%%%%%%

\subsubsection{Spectrum of the Model and Group Theoretical Interpretation}

%%%%%%%%%%%%%%%%%%%%%%%%%%%

Although, as we have seen, the linearized field equations for the
HS fields arising from the ${\cF}$ constraint exhibit a mixing
phenomenon, the physical degrees of freedom and the group
theoretical interpretation of the resulting spectrum can be
deduced from the mass-shell conditions \eq{mass} for the scalar
field $\phi$ and the Weyl tensors. This is due to the fact that
all local degrees of freedom enter the system via the $0$-form
sources. Thus, \emph{the mode expansions of a gauge-fixed
symmetric tensor and its Weyl tensor are based on the same
lowest-weight space $D(E_0;S_0)$}, and simply differ in the
embedding conditions for the irreducible Lorentz representation,
that we label by $J$.

In order to determine the lowest-weight spaces that arise in the
spectrum, one can thus treat $AdS_D$ as the coset space
$SO(D-1,2)/SO(D-1,1)$ and perform  a standard harmonic analysis
\cite{Salam:1981xd} of \eq{mass}, expanding $C^{\{s,s\}}$ in terms
of all $SO(D-1,2)$ irreps that contain the $SO(D-1,1)$ irrep
$\{s,s\}$. It follows that these irreps have the lowest weight
$\{E_0,S_0\}$, where $S_0\equiv \{s_1,s_2\}$ (recall that we are
suppressing the zeros, as usual) with $E_0 \ge s\ge s_1\ge s\ge
s_2 \ge 0 $. Taking into account the Bianchi identity satisfied by
$\Phi^{(s,s)}$, one can then show that $s_2=0$ \cite{s3}, and
therefore
 \be S_0=\{s\}\ .\ee
Next, one can use the standard formula that relates the Laplacian
$\nabla^2$ on $AdS_D$ acting on the $SO(D-1,2)$ irreps described
above to the difference of the second order Casimir operators for
the $SO(D-1,2)$ irrep $\{E_0, \{s\}\}$ and the $SO(D-1,1)$ irrep
$\{s,s\}$, thus obtaining the characteristic equation
 \be \frac14 \Big(~C_2[SO(D-1,2)|E_0;\{s\}]~-~C_2[SO(D-1,1)|\{s,s\}]~\Big)~+~
 \frac12\, (s+D-3)\ =\ 0\ .
 \ee
The well-known formula for the second-order Casimir operators
involved here then leads to
 \be
 \frac14\Big(\left[ E_0(E_0-D+1)+s(s+D-3)\right] -\left[
 2s(s+D-3)\right]
 \Big)~+~ \frac12\,(s+D-3)\ =\ 0\
,\ee
with the end result that
 \be
 E_0\ =\ {D-1\over 2}\pm \left(s+{D-5\over 2}\right)\
 =\ \mx{\{}{l}{s+D-3\\[5pt]2-s}{.}\ .
 \label{E0}
 \ee

The root $E_0=2-s$ is ruled out by unitarity, except for $D=4$ and
$s=0$, when both $E_0=2$ and $E_0=1$ are allowed. These two values
correspond to Neumann and Dirichlet boundary conditions on the
scalar field, respectively. Hence, the spectrum for $D \geq 5$ is
given by
\be {\mathcal S}_D\ =\ \bigoplus_{s=0,2,4,...}
D(s+D-3,\{s\}_{D-1})\ . \label{spectrum}\ee
while the theory in $D=4$ admits two possible spectra, namely
${\cal S}_4$ and
\be {\mathcal S}'_4\ =\ D(2,0)\ \oplus\quad\bigoplus_{s=2,4,...}
D(s+1,s)\ . \label{spectrum2}\ee
In $D=3$ the twisted-adjoint representation is one-dimensional,
and hence \cite{misha,s3,ps}
\be{\cal S}_3\ =\ R \ .\ee

As discussed in Section \ref{sec:t0}, the spectrum \eq{spectrum}
fills indeed a unitary and irreducible representation of
$ho_0(D-1,2)$ isomorphic to the symmetric product of two scalar
singletons \cite{misha,ps}. The alternative 4D spectrum ${\mathcal
S}'_4$ is also a unitary and irreducible representation of
$ho_0(3,2)$, given by the anti-symmetric product of two spinor
singletons $D(1,\frac12)$. These arise most directly in the spinor
oscillator realization of $ho_0(3,2)$, often referred to as
$hs(4)$ (see \eq{iso}).

%%%%%%%%%%%%%%%%%%%%%%%%%%%%%%%%%%%%%%%%%%%%%%%%%%%%%%%%%%%%%%%%%%

\subsection{Compensator Form of the Linearized Gauge-Field Equations}

%%%%%%%%%%%%%%%%%%%%%%%%%%%%%%%%%%%%%%%%%%%%%%%%%%%%%%%%%%%%%%%%

Leaving aside the mixing problem, and considering for simplicity
the flat limit, one is thus faced with the linearized curvature
constraints ($k=0,\dots,s-1$)\footnote{In this section we use the
notation in which $a(k)$ denotes a symmetric set of $k$ Lorentz
vector indices. }
\bea \cF^{(s-1,k)}_{\mu\nu,a(s-1),b(k)}&\equiv& 2\partial_{[\mu}
W^{(s-1,k)}_{\n],a(s-1),b(k)} + 2c_{s,k}
W^{(s-1,k+1)}_{[\mu|,a(s-1),|\nu]b(k)}\nn\\&=& \delta_{k,s-1}
C^{\{s,s\}}_{[\mu|a(s-1),|\nu]b(s-1)}\ ,\label{eq:fr6}\eea
where $c_{s,k}$ is a constant, whose precise value is
inconsequential for our purposes here, and the gauge fields
$W_\mu^{(s-1,k)}$ are traceful, while the 0-form $C^{\{s,s\}}$ on
the right-hand side is the \emph{traceless} spin-$s$ Weyl tensor.
The trace of \eq{eq:fr6} in a pair of internal indices generates a
homogeneous set of cohomological equations of the type considered
by Dubois-Violette and Henneaux \cite{dvh}, and the analysis that
follows is in fact a combination of the strongly projected
Vasiliev equations with the results of Bekaert and Boulanger
\cite{bb} on the link between the Freedman-de Wit connections and
the compensator equations of
\cite{Francia:2002aa,Sagnotti:2003qa}. Let us begin by discussing
some preliminaries and then turn to the cases of spin 3 and 4.
These exhibit all the essential features of the general case, that
will be discussed in \cite{s3}.

The scheme for eliminating auxiliary fields parallels the
discussion of the off-shell case. Thus, after fixing the
St\"uckelberg-like shift symmetries, one is led to identify the
independent spin-$s$ gauge field, a totally symmetric rank-$s$
tensor $\phi_{a(s)}\equiv \phi_{a_1\dots a_s}$, via
\be W^{(s-1,0)}_{\mu,a(s-1)}\ =\ \f_{a(s-1)\mu}\ ,\ee
where the hooked Young projection can be eliminated by a
gauge-fixing condition. The constraints on
$\cF^{(s-1,k-1)}_{\mu\nu}$, for $k=1,...,s-1$, determine the
\emph{Freedman-de Wit connections}, or generalized Christoffel
symbols
\be W^{(s-1,k)}_{\mu,a(s-1),b(k)}\ \equiv\ \c_{s,k}\
\partial^{k} \underbrace{{}_{b(k)}
\f_{a(s-1)}}_{(s-1,k)}{}_{\mu}\ , \label{eq:fr7}\ee
where the subscript $(s-1,k)$ defines the Young projection for the
right-hand side and the $\c_{s,k}$ are constants whose actual
values are immaterial for the current discussion. The remaining
gauge transformations are then, effectively,
\be \d_{\L}\f_{a(s)} \ = \ s\, \partial_{(a_1}\L_{a_2 ... a_{s})}\
,\label{eq:fr4}\ee
and are of course accompanied by gauge-preserving shift parameters
$\e^{(s-1,k)}$. It is important to note that the gauge-fixed frame
fields $W^{(s-1,k)}$ ($k=0,1,...,s-1)$ are \emph{irreducible}
$SL(D)$ tensors of type $(s,k)$, since their $(s-1,k+1)$ and
$(s-1,k,1)$ projections vanish as a result of too many
antisymmetrizations. On the other hand, the traces
$W^{(s-1,k)}_{\mu,a(s-1),b(k-2)c}{}^c$ for $k=2,\dots,s-1$ are
\emph{reducible} $SL(D)$ tensors, comprising $(s,k-2)$ and
$(s-1,k-1)$ projections (the remaining $(s-1,k-2,1)$ projection
again vanishes due to too many antisymmetrizations). In
particular, the $(s-1,k-1)$ part does not contain the
d'Alembertian and takes the form
\be
\underbrace{\hspace{-10pt}W^{(s-1,k)}_{\mu,\a(s-1),b(k-2)}}_{(s-1,k-1)}\
=\
\underbrace{\hspace{-5pt}\partial^{k-1}_{b(k-1)}j_{a(s-1)}}_{(s-1,k-1)}\
,\label{eq:fr9}\ee
where the constructs $j_{a(s-1)}$ are linear combinations of
$\partial\cdot \f_{a(s-1)}$ and $\partial_a\f'_{a(s-2)}$. The
above observations imply that the trace of \eq{eq:fr6} in a pair
of internal indices
\be  2\partial_{[\mu} W^{(s-1,k)}_{\n],a(s-1),b(k-2)c}{}^c +
2c_{s,k} W^{(s-1,k+1)}_{[\mu|,a(s-1),|\nu]b(k-2)c}{}^c\ =\ 0\
,\qquad k=2,\dots,s-1\ , \label{eq:fr61}\ee
results in a homogenous cohomological problem for
$W^{(s-1,k)}_{\mu,a(s-1),b(k-2)c}{}^c$ for $k=2,\dots,s-1$, whose
integration gives rise to the compensators as ``exact'' forms.

For the sake of clarity, let us begin by examining the spin-3
case. Here the relevant trace part, given by
\be W^{(2,2)}_{\mu,ab,c}{}^c\ =\ {\c_{3,2}\over 3}\, {\cal
K}_{\mu,ab}\ ,\qquad {\rm where}  \quad {\cal K}_{\mu,ab}\ = \
\Box \f_{ab\mu}-2\partial_{(a}\partial\cdot\f_{b)\mu}
+\partial_a\partial_b\f'_\mu\ , \ee
and where $\phi'$ denotes the trace of $\phi$ and $\c_{3,2}$ is a
constant whose actual value plays no role in our argument, obeys
the constraint
\be \partial_{[\mu} W^{(2,2)}_{\n],ab,c}{}^c\ =\ 0\ .\ee
Hence, ${\cal K}$ can be related to a symmetric rank-$2$ tensor:
\be {\cal K}_{\mu,ab}\ =\
\partial_\mu\b_{ab}\ , \label{eq:s32}\ee
where $\b_{a,b}$ is constrained since the $(2,1)$ projection of
\eq{eq:s32}, equivalent to anti-symmetrizing on $\mu$ and $a$,
implies that
\be \partial_{[\mu}(\partial\cdot \f_{a]b}-\partial_{a]}\f'_b)\ =\
\partial_{[\mu}\b_{a]b}\ ,\label{eq:s33}\ee
whose solution reads
\be \b_{ab}\ =\ \partial\cdot \f_{ab}-2\partial_{(a}\f'_{b)}\ + \
3
\partial_a\partial_b\a\ .\label{eq:s34}\ee
\emph{Notice that the last term is a homogeneous solution
parameterized by an unconstrained scalar function $\a$ and
involves two derivatives}. Plugging this expression for $\b_{ab}$
into the $(3,0)$ projection of \eq{eq:s32} finally gives
\be \cF_{abc}\ \equiv \ \Box
\f_{abc}-3\partial_{(a}\partial\cdot\f_{bc)}+
3\partial_{(a}\partial_b\f'_{c)}\ =\ 3 \partial^3_{abc}\a\
,\label{eq:s35}\ee
the flat-space version of the spin-$3$ compensator equation
\eq{eq:fr1}.

The origin of the triple gradient of the compensator can be made
more transparent by an alternative derivation, that has also the
virtue of generalizing more simply to higher spins. To this end,
one can analyze the consequences of the invariance under the gauge
transformation
\be \d_\L\f_{abc}\ =\ 3\partial_{(a}\L_{bc)}\ ,\ee
that affects the left-hand side of \eq{eq:s32} according to
\be \d_\L{\cal K}_{\mu,ab}\ =\ \partial_\mu \widetilde \L_{ab}\
,\qquad \widetilde{\L}_{ab}\ =\ \Box
\L_{ab}-2\partial_{(a}\partial\cdot\L_{b)}+
\partial^2_{ab}\L'\ .\ee
Gauge invariance thus implies that the homogeneous solution
$\beta_{ab}$ must transform as
\be \delta_\L\b_{ab}\ =\ \widetilde{\L}_{ab}\ .\ee

From $\d_\L\partial\cdot\f_{ab}=\Box
\L_{ab}+2\partial_{(a}\partial\cdot\L_{b)}$ and
$\d_\L\partial_{(a}\f'_{b)}=
\partial_{a}\partial_{b}\L'+2\partial_{(a}\partial\cdot\L_{b)}$, one can see
that $\b_{ab}$ is a linear combination of $\partial\cdot\f_{ab}$,
$\partial_{(a}\f'_{b)}$ and of a term $\partial^2_{ab}\a$, with
$\a$ an independent field transforming as
\be \d_\L\a\ =\ \L' \ . \ee
This in its turn implies that the symmetric part of \eq{eq:s32} is
a gauge covariant equation built from $\Box\f_{abc}$,
$\partial_{(a}\partial\cdot\f_{bc)}$, $\partial^2_{(ab}\f'_{c)}$
and $\partial^3_{abc}\a$, and uniquely leads to the compensator
equation \eq{eq:s35}.

The direct integration for spin $4$ is slightly more involved,
since two traced curvature constraints,
\bea \partial_{[\mu}W^{(3,3)}_{\nu],a(3),bc}{}^c&=&0\ ,\\
\partial_{[\mu}W^{(3,2)}_{\nu],a(3),c}{}^c\ +\ c_{4,2}\,W^{(3,3)}_{[\mu|,a(3),|\nu] c}{}^c&=&0\ ,\eea
have to be dealt with in order to obtain the compensator equation.
The first constraint implies that
\be W^{(3,3)}_{\m,a(3),bc}{}^c\ =\
\partial_\mu\b^{(3,1)}_{a(3),b}\ ,\label{eq:s41}\ee
where the explicit form of the Freedman-de Wit connection is
\be {4\over \gamma_{4,3}}\,W^{(3,3)}_{\m,a(3),bc}{}^c
\,=\,\partial_b\Box \f_{a(3)\m} \!-\! 2\partial_a\Box
\f_{a(2)b\mu}\!-\! \partial^2_{ab}
\partial\cdot\f_{a(2)\mu}\!-\! \partial^3_{a(2)b}\f'_{b\mu}
\!-\! 2\partial^2_{a(2)}\partial
\cdot\f_{ab\mu}+\partial^3_{a(3)}\f'_{b\mu}\ , \label{eq:s42}\ee
with $\c_{4,3}$ a constant whose actual value plays no role in our
argument. The $(3,2)$ projection does not contain any
d'Alembertians, and can thus be identified with the double
gradient of ${\cal J}_{a(3)}$, a construct of $\partial \cdot
\f_{a(3)}$ and $\partial_a\f'_{a(2)}$. It can be integrated once,
with the result that
\be \b^{(3,1)}_{a(3),b}\ =\ \underbrace{\partial_{b}{\cal
J}_{a(3)}}_{(3,1)}\ + \ \underbrace{\partial^3_{a(3)}
\a_{b}}_{(3,1)}\ ,\label{eq:s43}\ee
where the last term, with $\a_b$ an unconstrained vector field, is
the general solution of the homogeneous equation, \emph{i.e.} its
cohomologically exact part  in the language of Dubois-Violette and
Henneaux \cite{dvh}. Eq.~\eq{eq:s43} implies that the second
curvature constraint can be written in the form
\be \partial_{[\mu}\,W^{(3,2)}_{\n],a(3),c}{}^c\ +\,c_{4,2}
\,\partial_{[\m|}\b^{(3,1)}_{a(3),|\n]}\ =\ 0\ ,\ee
and integrating this equation one finds
\be W^{(3,2)}_{\mu,a(3),c}{}^c\ +\ c_{4,2}\,\b^{(3,1)}_{a(3),\m} \
=\ \partial_\mu \b^{(3)}_{a(3)}\ ,\label{eq:s44}\ee
where the homogeneous term $\beta^{(3)}$ is constrained,  since a
derivative can be pulled out from the $(3,1)$ projection of the
left-hand side. Thus, using
\be \underbrace{\partial^3_{a(3)}\a^{\phantom{3}}_{b}}_{(3,1)}\ =\
-\underbrace{\partial^{\phantom{1}}_{b}
\partial^2_{a(2)}\a^{\phantom{1}}_{a}}_{(3,1)}\
,\ee
one finds that
\be\b^{(3)}_{a(3)}\ =\ \widetilde {\cal J}_{a(3)}-
\,c_{4,2}\,\partial^2_{a(2)}\a_a\ ,\ee
where $\widetilde {\cal J}_{a(3)}$ is another construct of the
first derivatives of $\f$, and the homogeneous solution has been
absorbed into a shift of $\a_a$ by a gradient. Finally, the
$(4,0)$ projection of \eq{eq:s44} reads
\be W^{(3,2)}_{a,a(3),c}{}^c\ =\ \partial^{\phantom{1}}_a
\widetilde{\cal
J}_{a(3)}~-~c_{4,2}\partial^3_{a(3)}\a^{\phantom{1}}_a\ ,\ee
and gauge invariance implies that, up to an overall rescaling of
$\alpha_a$, this is the flat-space compensator equation
\eq{eq:fr1} for spin 4.

In summary, a careful analysis of the Vasiliev equations shows
that, if the gauge fields are left free to fluctuate and adjust
themselves to constrained Weyl 0-form sources, one is led, via the
results of \cite{bb}, to the compensator equations of
\cite{Francia:2002aa,Sagnotti:2003qa} rather than to the
conventional Fronsdal formulation.

%%%%%%%%%%%%%%%%%%%%%%%%%%%%%%%%%%%%%%%%%%%%%%%%%%%%%%%%%%%%%%%%%%%%%%

\section{TOWARDS A FINITE CURVATURE EXPANSION}\label{sec:proj}

%%%%%%%%%%%%%%%%%%%%%%%%%%%%%%%%%%%%%%%%%%%%%%%%%%%%%%%%%%%%%%%%%%%%%

In this section we state the basic problem one is confronted with
in the perturbative analysis of the strong $Sp(2,R)$-invariance
condition, and discuss how finite results could be extracted from
the non-linear Vasiliev equations in this setting. These
observations rest on properties of the $\star$-products of the
singular projector and other related non-polynomial objects, that
we shall illustrate with reference to a simpler but similar case,
the strong $U(1)$ condition that plays a role in the 5D
vector-like construction \cite{5d,mishacubic} based on spinor
oscillators \cite{gunaydin}. Whereas these novel possibilities
give a concrete hope that a finite construction be at reach, a
final word on the issue can not forego a better understanding of
the interactions in the actual $Sp(2,R)$ setting. We intend to
return to these points in \cite{s3}.

%%%%%%%%%%%%%%%%%%%%%%%%%%%%%%%%%%%%%%%%%%%%%%%%

\subsection{On the Structure of the Interactions}
\label{sec:dress}

%%%%%%%%%%%%%%%%%%%%%%%%%%%%%%%%%%%%%%%%%%%%%%%%

In the previous section, we found that the strong $Sp(2,R)$
condition \eq{kijproj} led to a \emph{linearized} zero-form master
field $\Phi$ with an expansion involving non-polynomial dressing
functions $F(N;K^2)$, according to \eq{rellon}. In particular, the
master-field projector $M$, introduced in \eq{MstarR} and such
that $\Phi=M\star C$,
\be M(K^2)\ =\ F(0;K^2)\ ,\ee
is the dressing function of the scalar field $\phi$ component of
$\Phi$. The dressing functions are proportional to $J_\nu
(\xi)/(\xi/2)^\nu$, where $J_\nu$ is a Bessel function of order
$\nu= (N+D-2)/2$ and of argument $\xi=4 \sqrt{K^2}$, and hence
satisfy the Laplace-type differential equation
\be \left( \xi \, \frac{d^2}{d\xi^2} + (2 \nu + 1) \
\frac{d}{d\xi} + \xi \right) \, F(N;\xi^2/16) \ = \ 0 \ .
\label{ode2}\ee
The solution of this equation that is analytic at the origin can
be given the real integral representation
\be F(N;K^2) \ = \ {\cal N}_\nu\, \int_{-1}^1 ds \ (1-s^2)^{\nu -
\frac12}~\cos\left(4 \sqrt{K^2} \; s\right)\ , \label{Mint}\ee
where the normalization, fixed by $F(N;0)=1$, is given by
\be {\cal N}_\nu\ =\ \frac{1}{B\left(\nu+\frac12,\frac12\right)}\
=\ \frac{\Gamma(\nu+1)}{\Gamma\left(\nu+\frac{1}{2}\right)
\Gamma\left(\frac12\right)}\ , \label{normalize} \ee
with $B$ the Euler $B$-function. Alternatively, in terms of the
variable $z = K^2$, the differential equation takes the form
\be \left( \frac{z}{4} \, \frac{d^2}{dz^2} + \frac{ \nu + 1}{4} \
\frac{d}{dz} + 1 \right) \, F(N;z) \ = \ 0 \ , \label{ode1}\ee
which is also of the Laplace type, and leads directly to the
Cauchy integral representation
\be F(N;K^2)\ =\ \Gamma(\nu+1)\oint_\gamma \frac{dt}{2 \pi i}
\frac{\exp\left( t - \frac{4K^2}{t} \right)}{t^{\nu+1}} \ , \ee
where the Hankel contour $\gamma$ encircles the negative real axis
and the origin.

In building the perturbative expansion of the strongly projected
zero form, one also encounters an additional non-polynomial
object, that we shall denote by $G(K^2)$ and plays the role of a
$\star$-\emph{inverse} of the $Sp(2,R)$ Casimir operator. It is
the solution of
\be K^I\star K_I\star G\star H\ =\ H\ ,\label{KKG}\ee
where $H$ belongs to a certain class of functions not containing
$M(K^2)$ \cite{s3}. This element instead obeys
\be K^I\star K_I\star G\star M\ =\ 0\ ,\label{KKGM}\ee
in accordance with the associativity of the $\star$-product algebra.

Dealing with the above non-polynomial operators, all of which can
be given contour-integral representations, requires pushing the
$\star$-product algebra beyond its standard range of
applicability. The integral representations can be turned into
forms containing exponents linear in $K_I$, suitable for
$\star$-product compositions, at the price of further parametric
integrals, but one then discovers the presence of a
one-dimensional \emph{curve of singularities} in the parametric
planes, leading to an actual \emph{divergence} in $M\star M$
\cite{misha}.

Let us consider in more detail the structure of the interactions.
At the $n$-th order they take the form
\be \widehat {\cal O}_1\star M\star \widehat{\cal O}_2\star
M\star\widehat{\cal O}_3\star \cdots \star M\star\widehat{\cal
O}_{n+1}\ ,\label{struct}\ee
where the $\widehat {\cal O}_s$, $s=1,\dots,n+1$ are built from
components of the master fields $A_\mu$ and $C$ defined in
\eq{icF}, the element $\kappa(t)$ defined in \eq{kappat}, and the
$\star$-inverse functions $G(K^2)$. One possible strategy for
evaluating this expression would be to group the projectors
together, rewriting it as
\be  \widehat {\cal O}_1\star \underbrace{M\star M\star\cdots
\star M}_{\mbox{$n$ times}}\star [\widehat{\cal O}_2]_{0}\star
[\widehat{\cal O}_3]_0\star \cdots \star[{\cal
O}_{n}]_0\star\widehat{\cal O}_{n+1}\ ,\label{rearr}\ee
where $[\widehat{\cal O}]_0$ denotes the $Sp(2,R)$-singlet, or
neutral, projection\footnote{In general, any polynomial in
oscillators can be expressed as a finite sum of $(2 k + 1)$-plets.
The neutral part is the singlet, $k=0$, component. A properly
refined form of eq.~\eq{rearr} applies also to more general
structures arising in the $\Phi$-expansion, where the inserted
operators can carry a number of doublet indices.} of
$\widehat{\cal O}$. This procedure clearly leads to divergent
compositions that appear to spoil the curvature expansion.
However, \eq{rearr} is only an intermediate step in the evaluation
of \eq{struct}, and the singularities may well cancel in the final
Weyl-ordered form of \eq{struct}. Therefore, one should exploit
the associativity of the $\star$-product algebra to find a
strategy for evaluating \eq{struct} that may avoid intermediate
singular expressions altogether. To achieve this, one may first
$\star$-multiply the projectors with adjacent operators
$\widehat{\cal O}_s$, and then consider further compositions of
the resulting constructs. It is quite possible that this
rearrangement of the order of compositions result in interactions
that are actually completely free of divergences, as we shall
discuss further below.

In principle, one may instead consider a modified curvature
expansion scheme, based on an additional $Sp(2,R)$ projector,
which we shall denote by $\Delta(K^2)$, that is of distributional
nature and has finite compositions with both the analytic
projector and itself \cite{s3}. We shall return to a somewhat more
detailed discussion of possible expansion schemes in Section
\ref{sec:finite}, while we next turn to the analysis of a simpler
$U(1)$ analog.

%%%%%%%%%%%%%%%%%%%%%%%%%%%%%%%%%%%%%%%%%%%%%%%%%%%%%%%%%%%%%%

\subsection{A Simpler $U(1)$ Analog}

%%%%%%%%%%%%%%%%%%%%%%%%%%%%%%%%%%%%%%%%%%%%%%%%%%%%%%%%%%%%%%%

In this section we examine a strong $U(1)$ projection that is
simpler than the actual $Sp(2,R)$ case but exhibits similar
features. The key simplification is that the corresponding
non-polynomial objects admit one-dimensional parametric
representations, directly suitable for $\star$-product
compositions, that only contain \emph{isolated singularities} in
the parameter planes. Drawing on these results, in Section
\ref{sec:finite} we shall finally discuss two possible
prescriptions for the curvature expansion.

%%%%%%%%%%%%%%%%%%%%%%%%%%%%%%%%%%%%%%%%%%%%%%%%%%%%%%%%%%%%%%%

\subsubsection{Analytic $U(1)$ Projector}

%%%%%%%%%%%%%%%%%%%%%%%%%%%%%%%%%%%%%%%%%%%%%%%%%%%%%%%%%%%%%%%%%%%%%

The $Sp(2,R)$ dressing functions $F(N;K^2)$ are closely related
to others occurring in the 5D and 7D constructions of \cite{5d}
and \cite{7d}. These are based on bosonic Dirac-spinor oscillators
\cite{gunaydin} $y_\alpha$ and $\bar{y}^\beta$ obeying
\be y_\alpha \star \bar{y}^\beta \ = \ y_\alpha \bar{y}^\beta +
\delta_\alpha^\beta \ ,\ee
and providing realizations of the minimal bosonic higher-spin
algebras $ho_0(4,2)$ and $ho_0(6,2)$, at times referred to as
$hs(2,2)$ and $hs(8^*)$, via on-shell master fields subject to
$U(1)$ and $SU(2)$ conditions, respectively. In particular, the 5D
$U(1)$ generator is given by \cite{5d}
\be x\  =\   \bar{y} y \ .  \ee
One can show that for $2n$-dimensional Dirac spinor oscillators
\be x\star f(x)\ =\ \left(x\ - \  2\, n\, \frac{d}{dx} \ - \ x\,
\frac{d^2}{dx^2}  \right) f(x)\ ,\label{xfx}\ee
where $n=2$ in $D=5$. The strong $U(1)$-invariance condition
defining the 5D Weyl zero form \cite{5d} can be formally solved
introducing a projector $m(x)$ analytic at the origin and such
that \cite{mishacubic}
\be x \star m(x) \ = \ 0 \ . \ee
The analytic projector $m$ belongs to a class of $U(1)$ dressing
functions $f(N;x)$, that arise in the component-field expansion of
the linearized Weyl master zero form \cite{5d}, and which obey
\be \left( x \, \frac{d^2}{dx^2} \ + \ (2\nu+1)\, \frac{d}{dx} \ -
\ x \right) f(N;x) \ = \ 0 \ ,\label{lap2}\ee
with $\nu=n+N-\frac12$, so that
\be m(x)\ =\ f(0;x)\ .\label{mfx}\ee
The solution of \eq{lap2} that is analytic at the origin can be
expressed in terms of the modified Bessel function of index $\nu$
and argument $x$ as $I_{\nu}(x)/x^\nu$, and admits the real
integral representation \cite{ww}
\be f(N;x)\ =\ {\cal
N}_{\nu}~\int_{-1}^{1}~ds~(1-s^2)^{\nu-\frac12}~e^{sx}\ ,\ee
where the normalization ${\cal N}_\nu$, determined by the
condition that $f(N;0)=1$, is as in \eq{normalize}.

In order to compose $m$ with other objects, it is useful to recall
the $\star$-product formula
\be e^{sx} \star e^{s'x} \ = \ \frac{e^{ t(s,s')x}}{(1+s s')^{2n}} \ ,
\quad \mbox{with  \ } t(s,s') \ = \ \frac{s+s'}{1 + s s'} \ ,
\label{tss} \ee
that holds for arbitrary values of $s,s'\in\mathbf{C}$ such that
$1+ss'\neq 0$, and to note that the projective transformation
$t(s,s')$ has the form of the relativistic addition of velocities,
so that $t(\pm 1,s')=t(s,\pm 1)=\pm 1$. This can be used to show
that, for any complex number $\lambda\neq \pm 1$,
\be e^{\lambda x}\star m(x)\ =\ \frac{1}{(1-\lambda^2)^{n}}\ m(x) \
,\label{eq:l1}\ee
and hence that the $\star$-product of a pair of analytic
projectors results in the singular expression
\be m(x) ~\star ~m(x) \ = \ {\cal N}~ m(x)~\int_{-1}^1 {ds\over 1-s^2}\
 \ , \label{inf}\ee
where ${\cal N}\equiv{\cal N}_{n-\frac12}$. The logarithmic nature of
the singularity is consistent with power counting in the
double-integral expression
\be m(x)~\star~ m(x)\ =\ {\cal
N}^2\int_{-1}^1~ds~\int_{-1}^1~ds'~{(1-s^2)^{n-1}(1-s^{\prime
2})^{n-1}\over (1+ss')^{2n}}~e^{t(s,s')x}\ee
where the integrand diverges as $\epsilon^{-2}$ as $s\sim
1-\epsilon$ and $s'\sim -1+\epsilon$.

As discussed in Section \ref{sec:dress}, the above singularity
need not arise in the curvature expansion of a HS theory based on
the spinor oscillators. Indeed, if one first composes the $m$'s
with other operators, expands in terms of dressing functions, and
continues by composing these, one encounters
\be f(N;x)~\star~ f(N';x)\ =\ {\cal N}_{\nu}~{\cal N}_{\nu'}
\sum_{k=0}^{N'}{2N'\choose2k}\, B\left(k+\frac12,N+N'\right)\,
I_{N,N';k}(x)\ ,\label{fPQ}\ee
with $\nu=n+N-\frac12$, $\nu'=n+N'-\frac12$ and
\be I_{N,N';k}(x)\ =\  \int_{-1}^1 ~dt~ t^{2k}
(1-t^2)^{\nu'-\frac12} \, {}_2
F_1\left(k+\frac12,2N';N+N'+k+\frac12;t^2\right) ~e^{tx}\ , \ee
where ${}_2 F_1$ is the hypergeometric function. This type of
expression is free of divergences for $N+N'\geq 1$, while it is
apparently equal to the singular expression ${\cal N}\,
\Gamma(0)\, m(x)$ if $N=N'=0$, but in fact, using ${}_2
F_1(0,\beta;\gamma;z)=1$, one can see that
\be f(N;x)~\star~m(x)\ =\ {\cal N}_\nu~B\left(N,\frac12\right)~m(x)\ .
\ee
More generally, multiple compositions $f(N_1;x)\star\cdots \star
f(N_n;x)$ are finite for $N_1+\cdots+N_n\geq 1$, as can be seen by
power counting. Therefore, the finiteness of the interactions is
guaranteed provided the case $N_1=\cdots=N_n=0$ never presents
itself, a relatively mild
condition that could well hold for the interactions
\footnote{This corresponds, roughly speaking, to
constraints on scalar-field \emph{non-derivative} self interactions,
which might be related to the fact that all such
couplings actually vanish in the 4D spinor-formulation
as found in \cite{Sezgin:2003pt}.}.

It is interesting to examine
more closely the nature of the singularity in \eq{inf}.\\[10pt]
%
%%%%%%%%%%%%%%%%%%%%%%%%%%%%%%%
%
\begin{centerline}{\it Problems with Regularization of the
Singular Projector}
\\[10pt]\end{centerline}
%
%%%%%%%%%%%%%%%%%%%%%%%%%%%%%%%
%
The singularities of \eq{inf} can be removed by a cutoff
procedure, at the price of violating $U(1)$ invariance. For
instance, given the regularization
\be m_\epsilon(x) \ = \ {\cal N} \int_{-1+\epsilon}^{1-\epsilon} ds \
e^{s x} \ (1-s^2)^{n-1} \ , \ee
one can use \eq{xfx} to show that
\be x\star m_\epsilon(x)\ =\  2~{\cal N}\,
[1-(1-\epsilon)^2]^{n}~\sinh[(1-\e)x]\ ,\ee
where $\sinh[(1-\e)x]$ belongs to the ideal, since $\sinh[\lambda
x] \, \star\, m(x)=0$ for all $\lambda$ . However, while the
resulting violation may naively appear to be small, it is actually
sizeable, since $\sinh[(1-\e)x]$ and $m_\epsilon(x)$ have a very
singular composition, with the end result that an \emph{anomalous
finite violation} of $U(1)$ invariance can emerge in more
complicated expressions, so that for instance
\be x\star m_\epsilon(x) \star m_\epsilon(x) \ =\ {\cal N}^2~\int_0^1
ds~ (1-s^2)^{n-1} \sinh (sx)~ +~ \mbox{evanescent}\ .\ee
This anomaly can equivalently be computed first composing
\be m_\epsilon(x) \star m_\epsilon(x)~=~ {\cal N}^2
\int_0^{t(1-\epsilon,1-\epsilon)} du ~ \int_{-u}^u dt~ (1-t^2)^{n-1}~
e^{tx}\ ,\label{ano}\ee
and then expanding in $\epsilon$, which yields
\be m_\epsilon(x) \star m_\epsilon(x) \ =\ {\cal N} \log
\left(\frac{2-\epsilon}{\epsilon}\right)~ m(x) ~+~ A^{(2)}(x)~+~
\mbox{evanescent} \ ,\label{meme} \ee
where the divergent part is proportional to $m(x)$, while the
finite, anomalous, part
\be A^{(2)}(x)\ =\ {\cal N}^2 \int_0^1du~\log
\left(\frac{1+u}{1-u}\right)~(1-u^2)^{n-1}\sinh ux\ ,\ee
belongs to the ideal. One can verify that its composition with $x$
indeed agrees with the right-hand side of \eq{ano}, in compliance
with associativity. Continuing in this fashion, one would find
that higher products of $m_\epsilon(x)$ with itself keep producing
anomalous finite terms together with logarithmic singularities,
all of which are proportional to $m(x)$, together with tails of
evanescent terms that vanish like powers or powers times
logarithms when the cutoff is removed \cite{s3}:
\be \underbrace{m_\epsilon(x)\star \cdots
m_\epsilon(x)}_{\mbox{$p$ times}}\ =\ \sum_{l=-p+1}^{-1}
\left(\log\frac{1}{\epsilon}\right)^l~c^{(p)}_lm(x)~+~F^{(p)}(x)+\sum_{k\geq
1,l\geq 0} \epsilon^k\left(\log\frac 1{\epsilon}\right)^l
~E^{(p)}_{k,l}(x)\ ,\label{meme2}\ee
with $F^{(p)}(x)=c^{(0)} m(x)+A^{(p)}(x)$, where $A^{(p)}(x)$
represents the anomaly.

We would like to stress that the integral representations should
be treated with some care, as can be illustrated considering
\be m_\epsilon(x)\star m(x)\ =\ {\cal
N}~\log\left(\frac{2-\epsilon}{\epsilon}\right)~m(x)\ ,\ee
and its generalization
\be m(x)\star \underbrace{m_\epsilon(x)\star\cdots\star
m_\epsilon(x)}_{\mbox{$p$ times}} \ =\ \left({\cal
N}~\log\frac{2-\epsilon}{\epsilon}\right)^p~m(x)\ .\label{mme}\ee
Comparing \eq{mme} with \eq{meme}, it should be clear that one
should perform the parametric integrals prior to expanding in
$\epsilon$, since one would otherwise encounter ill-defined
compositions of ``bare'' $m$'s.

In order to appreciate the meaning of the anomaly, let us consider
a tentative Cartan integrable system\footnote{These considerations
would apply, in their spirit, to the 5D spinor construction, whose
completion into an integrable non-linear system is still to be
obtained.} on the direct product of spacetime with an internal
noncommutative $z$-space, containing a strongly $U(1)$-projected
0-form master field $\widehat \Phi$ with a formal perturbative
expansion in $m\star C$. At the $n$-th order, one could use the
$U(1)$ analog of the rearrangement in \eq{rearr} to bring all
projectors together, and then attempt to regulate the resulting divergent
$\star$-product compositions, for instance replacing each $m$ by an $m_\epsilon$.
This induces a violation of the constraints in
$z$-space. Therefore, in order to preserve the Cartan integrability
in spacetime, one could use \emph{unrestricted} $z$-expansions for the full
master fields, at the price of introducing ``spurious'' space-time
fields that one would have to remove by some form of consistent
truncation. To analyze this, one first observes that the structure
of the $\e$-dependence in \eq{meme2} implies that, the Cartan integrability
condition satisfied by the finite part of the regulated interactions can not contain
any contributions from singular times evanescent terms. Thus, the finite part
in itself constitutes a consistent set of interactions, albeit for all the
space-time fields, including the spurious ones.

In the absence of anomalies it would be consistent
to set to zero all the spurious fields, thus obtaining a well-defined
theory for the original space-time master fields. However, as the earlier
calculations show, these anomalies arise inevitably and jeopardize
these kinds of constructions, in that they must cancel in the final form
of the interactions, along lines similar to those discussed in Section 5.1
and indicated below \eq{fPQ}.

%%%%%%%%%%%%%%%%%%%%%%%%%%%%%%%%%%%%%%%%%%%%%%%%%%%%%%%%%%%%%%%

\subsubsection{Distributional $\star$-Inverse Function and
Normalizable Projector}\label{sec:green}

%%%%%%%%%%%%%%%%%%%%%%%%%%%%%%%%%%%%%%%%%%%%%%%%%%%%%%%%%%%%%%%

As we have seen, the $U(1)$ projection condition $x\star m(x)=0$
corresponds to the Laplace-type differential equation determined
by \eq{xfx}, which admits two solutions that are ordinary
functions of $x$, so that the singular projector $m(x)$ is the
unique one that is also analytic at $x=0$. Interestingly, there
also exist solutions that are distributions in $x$. Their Laplace
transforms involve $e^{sx}$ with an imaginary parameter $s$, which
improves their $\star$-product composition properties, and in
particular allows for a normalizable projector. A related
distribution, with similar properties, provides a $U(1)$ analog of
the $\star$-inverse of the $Sp(2,R)$ Casimir defined in \eq{KKG}.

To describe these objects, it is convenient to observe that, if
$\gamma$ is a path in the complex $s$-plane, the function
\be p_\gamma(x)\ =\ \int_\gamma ds\ (1-s^2)^{n-1}\ e^{sx}\ \ee
obeys
\be x\star p_\gamma\ =\ \left[(1-s^2)^{n}\
e^{sx}\right]_{\partial\gamma}\ .\label{pgamma}\ee
For instance, the projector $m(x)\sim I_\nu(x)/x^{\nu}$, with
$\nu=n-\frac12$, obtains taking for $\gamma$ the unit interval,
while including boundaries also at $\pm \infty$ yields in general
linear combinations of $I_\nu(x)/x^{\nu}$ and $K_\nu(x)/x^\nu$.
The latter is not analytic at $x=0$, however, and as such it would
not seem to play any role in the dressing of the linearized
zero-form master field. An additional possibility is provided by
the standard representation of Dirac's $\delta$ function,
\be \delta(x) \ = \ \int_{-\infty}^\infty \, \frac{dt}{2 \pi} \
e^{it x - \e t^2} \, \ee
with $\epsilon \to 0^+$, which suggests that boundaries at $\pm i\infty$
might give rise to projectors that are distributions in $x$.
To examine this more carefully, let
us consider
\be d(x)\ =\ \int_{- i \infty}^{i \infty} \, {ds\over 2\pi i}~
(1-s^2)^{n-1}~e^{xs+\epsilon s^2}\ .\label{Dint}\ee
Treating $d(x)$ as a distribution acting on test functions $f(x)$
such that $f^{(2k)}(0)$ fall off fast enough with $k$, one can
indeed expand it as
\be d(x)\ =\ \sum_{k=0}^{n-1} {n-1\choose k}(-1)^{k}\delta^{(2k)}(x)\
.\ee
One can now verify that $x\star d(x)=0$ holds in the sense that
$\left(x-2n\,\frac{d}{dx}-x\frac{d^2}{dx^2}\right)d(x)$ vanishes when
smeared against test functions,
\bea&& \int_{-\infty}^\infty dx \, f(x)\, (x\star
d(x))\label{calc}\\&&=\ \sum_{k=0}^{n-1}{{n-1\choose
k}}(-1)^{k}\left.\left[(xf(x))^{(2k)}+2n\,f^{(2k+1)}(x)-(xf(x))^{(2k+2)}
\right]\right|_{x=0}\ =\ 0\ .\nn\eea
We can now look more closely at the composition properties of $d(x)$,
beginning with
\be d(x) \star m(x)\ =\ {\cal N} \int_{-i \infty}^{i \infty} \,
{ds\over\pi i} \int_{-1}^1 ds'~{(1-s^2)^{n-1}(1-s'^2)^{n-1}\over
(1+ss')^{2n}}~e^{t(s,s')x+\epsilon s^2}\ ,\ee
that can be obtained from \eq{tss}. Notice that the denominator is
no longer singular, since $1+ss'\neq 0$ for $(s,s')\in iR\times
[-1,1]$, and after a projective change of integration variable,
$t(s,s')\rightarrow t$ at fixed $s$, one then finds
\be d(x) \star m(x) \ =\ {\cal N} \int_{-i \infty}^{i \infty}
{ds~e^{\epsilon s^2}\over\pi i(1-s^2)}~ \int_{\gamma_s}
dt~(1-t^2)^{n-1}~e^{tx}\ .\ee
The original real contour has been mapped to $\gamma_s$, a
circular arc from $t=-1$ to $t=+1$ crossing the imaginary axis at
$t=s$, but can be deformed back to the real interval $[-1,1]$
using Cauchy's theorem, so that
\be d(x) \star m(x) \ =\ m(x)~\left(\int_{-i \infty}^{i
\infty}\, {ds~e^{\epsilon s^2}\over\pi i(1-s^2)}\right)\ ,\ee
to be compared with \eq{inf}. The integral over $s$ is now
convergent, and is actually equal to $1$, and therefore
\be d(x) \star m(x) \ =\ m(x) \ .\label{Dmm}\ee
The iteration of this formula then yields
\be \underbrace{d(x) \star\cdots \star d(x) }_{\mbox{$p$
times}}~\star~ m(x) \ =\ m(x) \ .\label{Dmm2}\ee
This indicates that $d(x)$ is in fact \emph{normalizable}, that is
\cite{s3}
\be d(x) \star d(x) \ =\ d(x) \ ,\label{ddd}\ee
as can be seen by a direct computation using a prescription in
which the singularity at $ss'+1=0$ is avoided by small horizontal
displacements of the contour that are eventually sent to zero, as
in eqs.~\eq{Delta} and \eq{deltag} below.

Let us next turn to the $U(1)$ analog of the $\star$-inverse of the $Sp(2,R)$ Casimir.
This element, which we shall refer to as $g(x)$, is defined
by

\begin{itemize}

\item[i)] the inverse property
\be x\star g(x)~\star~ h(x)\ =\ h(x)\ ,\label{i}\ee
for an arbitrary ideal function $h(x)$. This milder form is the
counterpart of the condition for a distributional solution $d$ in
\eq{calc}.

\item[ii)] the twisted reality condition
\be (g(x))^\dagger\ =\ \sigma ~ g(-x)\ ,\label{ii}\ee
where $\sigma$ can be either $+1$ or $-1$;

\item[iii)] the orthogonality relation
\be m(x)~\star ~g(x)\ =\ 0\ .\label{iii}\ee

\end{itemize}

The two first conditions suffice to ensure that, given a linearized master
0-form obeying $x\star\Phi=0$ and $\tau(\Phi)=\Phi^\dagger=\pi(\Phi)$ \cite{5d},
one can construct a full master $0$-form $\widehat \Phi$ that satisfies
\begin{itemize}
\item[1)] a strong $U(1)$-invariance condition
\be \widehat x ~\star~\widehat\Phi\ =\ 0\ ,\ee
where $\widehat x$ is a full (non-linear) version of the $U(1)$ generator,
with a $\Phi$-expansion given by
\be \widehat x\ =\ x+\sum_{n=1}^\infty\widehat x_{(n)}\ ;\label{whx}\ee
\item[2)] the twisted $\tau$ and reality conditions
\be \tau(\widehat\Phi)\ =\ \widehat\Phi^\dagger\ =\ \pi(\widehat\Phi)\ .\ee
\end {itemize}

The perturbative expression for the full zero form is then given by
\be \widehat\Phi\ =\ (g(x)\star\widehat
x)^{-1}\star\Phi\star(\pi(\widehat x)\star g(x))^{-1}\ ,\ee
provided that \eq{i} applies to $g(x)\star x\star\widehat
\Phi_{(n)}$, in which case we note that the inverse elements in
the above formula can be expanded in a geometric series using
$g(x)\star\widehat x=1+\sum_{n=1}^\infty g\star \widehat x_{(n)}$.

To solve the conditions on $g(x)$, one can first use $\sinh(sx)\star
m(x)=0$ which follows from \eq{eq:l1}, to arrange that $g(x)\star
m(x)=0$ by writing $g(x)=\int_\gamma ds \sinh(sx) q(s)$. Using
\eq{pgamma}, a formal solution to the condition \eq{i} is then provided
by
\be g(x)\ =\ \left[\alpha
\int_{-i\infty}^0+(1-\alpha)\int_{-1}^0\right]
ds~(1-s^2)^{n-1}~\sinh(sx)~e^{\epsilon s^2}\ ,\ee
where $\epsilon\rightarrow 0^+$, $\alpha$ is a constant (to be fixed
below), and we have assumed that the boundary terms in \eq{pgamma} at
$\pm i\infty$ drop out when used in \eq{i}. These boundary terms play a
role, however, in verifying associativity in $x\star g(x)\star m(x)=0$.
Assuming that it is legitimate to set $\epsilon=0$ during these
manipulations, it follows that $x\star g(x)\star m(x)=(1-\alpha)m(x)$
which fixes $\alpha=1$.

In summary, we have found that distributions can be used to
construct a normalizable projector $d(x)$ as well as a
$\star$-inverse $g(x)$ of $x$, given by
\bea d(x) &=& \left(\int_{-i\infty+\eta}^0 +
\int_{0}^{i\infty-\eta}\right){ds\over \pi i}~(1-s^2)^{n-1}
~e^{sx+\epsilon s^2}\ ,\label{Delta}\\[10pt]g(x) &=& \frac12 \,\left(\int_{-i\infty+\eta}^0
+ \int_{i\infty-\eta}^0\right)ds~(1-s^2)^{n-1} ~e^{sx+\epsilon s^2}\ ,
\label{deltag}\eea
with $\epsilon,\eta \to 0^+$, where $\eta$ is a prescription for
avoiding singularities in $d(x)\star d(x)$ and $g(x)\star g(x)$.
These distributional objects can be expanded in terms of elementary
distributions using the standard result
\be \int_0^\infty dt \, e^{i x t} \ = \ \pi \, \delta(x) \ + \ i\,
PP\left( \frac{1}{x} \right) \ ,\label{PP} \ee
with $PP$ the principal part. Furthermore,
\be \tau(d(x))\ =\ \pi(\d(x))\ =\ \d(x)\ ,\qquad
\pi(g(x))\ =\ \tau(g(x))\ =\ -g(x)\ ,\ee
and both elements are hermitian, \emph{i.e.}
\be (d(x))^\dagger\ =\ d(x)\ ,\qquad (g(x))^\dagger\ =\
g(x)\ ,\ee
so that $\sigma=-1$ in (iii).

Finally, the generalization of $m(x)\star g(x)=0$ to arbitrary $U(1)$
dressing functions reads
\be f(N;x)~\star~g(x)\ =\ {i{\cal N}_\nu\over
2}~\sum_{k=0}^{N-1}(-1)^k{2N\choose 2k+1}B(k+1,N-k) I_{N;k}(x)\
,\label{fNg}\ee
where $\nu=n+N-\frac12$ and
\be I_{N;k}(x)\ =\ \int_{-1+i\delta}^{1+i\delta}
dt~t^{2k+1}~(1-t^2)^{\nu-\frac12}~{}_2F_1(k+1,2N;N+1;1-t^2)~e^{tx}\
,\ee
where $\delta\rightarrow 0^+$ is a prescription for how to encircle
poles at $t=0$, and we note that there are no inverse powers of $x$ in this
expression.

%%%%%%%%%%%%%%%%%%%%%%%%%%%%%%%%%%%%%%%%%%%%%%%%%%%%%%%%%%%%%%%%%%%%%

\subsection{Proposals for Finite Curvature Expansion Schemes}
\label{sec:finite}

%%%%%%%%%%%%%%%%%%%%%%%%%%%%%%%%%%%%%%%%%%%%%%%%%%%%%%%%%%%%%%%%%%%%%

We would like to conclude by summarizing our current understanding
of how a finite curvature expansion could be obtained from the
Vasiliev equations \eq{m4} supplemented with the strong $Sp(2,R)$
projection condition \eq{kijproj}. The material presently at our
disposal suggests two plausible such schemes, that we have partly
anticipated, and we shall refer to as minimal and modified. We
hope to report conclusively on the fate of these schemes in
\cite{s3}.

%%%%%%%%%%%%%%%%%%%%%%%%%%%%%%%%%%%%%%%%%%%%%%%%%%%%%%%%%%%%%%%%%%%%%

\subsubsection{Minimal Expansion Scheme} \label{sec:nat}

%%%%%%%%%%%%%%%%%%%%%%%%%%%%%%%%%%%%%%%%%%%%%%%%%%%%%%%%%%%%%%%%%%%%%

In this scheme, which is the most natural one, one solves the
internal constraints \eq{master1} and the strong projection
condition \eq{kijproj} by an expansion in the linearized Weyl zero
form $\Phi=M(K^2)\star C$. This object can be written using
dressing functions $f(N;K^2)$, as in \eq{rellon}. We anticipate
that the $\star$-products of the $Sp(2,R)$ dressing functions obey
an analog of \eq{fPQ}. The expansion also requires a
$\star$-inverse function $G(K^2)$ obeying \eq{KKG} and a suitable
set of boundary conditions, analogous to those discussed in the
$U(1)$ case. We expect this $\star$-inverse to be also
distributional and to satisfy an $Sp(2,R)$ analog of \eq{fNg}.

Let us consider an $n$-th order interaction of the form
\eq{struct}. Since the $\widehat{\cal O}_s$ are either arbitrary
polynomials or their $\star$-products with a $\star$-inverse
function, the products $M\star \widehat {\cal O}_s$ yield either
dressing functions or their products with the $\star$-inverse
function, respectively. Here we note that, anticipating an
$Sp(2,R)$ analog of \eq{fNg}, all parametric integrals associated
with (distributional) $\star$-inverse functions would not give
inverse powers of $K^2$. The resulting form of \eq{struct} could
be written as a multiple integral over a set of parameters $s_i\in
[-1,1]$ where the integrand contains factors
$(1-s_i^2)^{\nu_i-\frac12}$, where $\nu_i\geq (D-2)/2$, and a set
$\star$-products involving exponentials of the form $e^{s_i x}$.
These $\star$-products give rise to divergent powers of
$(1+s_is_j)$, and the idea is that all of these would be cancelled
by the remaining part of the integrand, resulting in a
well-defined interaction devoid of singularities.

As mentioned earlier, however, while the singularities in the
$U(1)$ case are isolated points in the parametric planes, in the
$Sp(2,R)$ case one has to deal with lines of singularities, and
this requires further attention before coming to a definite
conclusion.

%%%%%%%%%%%%%%%%%%%%%%%%%%%%%%%%%%%%%%%%%%%%%%%%%%%%%%%%%%%%%%%%%%%%%

\subsubsection{Modified Expansion Scheme} \label{sec:mod}

%%%%%%%%%%%%%%%%%%%%%%%%%%%%%%%%%%%%%%%%%%%%%%%%%%%%%%%%%%%%%%%%%%%%%

In this scheme, which is less natural than the previous one,
albeit still a logical possibility, one first solves the internal
constraints \eq{master1} and the strong projection condition
\eq{kijproj} by an expansion in $\Phi=\Delta(K^2)\star C$, where
the distributional $Sp(2,R)$ projector obeys $K_I\star
\Delta(K^2)=0$ and is given by \cite{s3}
\be \Delta(K^2)\ = \ \int_{-\infty}^{\infty} {ds\over 2 \pi}~
(1-s^2)^{\frac{D-3}2}~\cos (4\sqrt{K^2}s)\ . \label{Dint2} \ee

Starting from the above \emph{distributional master fields}, and
using the rearrangement in \eq{rearr} to bring all $\Delta$'s
together, one can define \emph{analytic master fields} $\widehat
A^\prime_\mu[A,\Phi]$, $\widehat A^\prime_i[\Phi]$ and $\widehat
\Phi^\prime[\Phi]$ by replacing the resulting single $\Delta$ by
an $M$. The so constructed analytical master fields obey
\bea \widehat F^\prime&\equiv& d\widehat A^\prime+\widehat A\star
\widehat A^\prime\ =\ \frac{i}2dZ^i\wedge dZ_i~\widehat
\Phi^\prime\star\kappa\ ,\label{mod5}\\[10pt] \widehat D \Phi^\prime&\equiv&
d\widehat\Phi^\prime+[\widehat A,\widehat \Phi^\prime]_\pi\ =\ 0\
, \label{mod4}\eea
where it is immaterial which of the two master fields in each of
the covariantizations that is primed, since the rearrangement in
\eq{rearr} implies that, schematically,
\be \widehat U^{(n)}\star \widehat V^{\prime(p)}\ =\ \widehat
U^{\prime(n)}\star\widehat V^{(p)}\ .\label{pc} \ee
Following steps similar to those outlined below eq. \eq{m6}
\cite{s3}, one can show the perturbative integrability of the
analytic constraints, eqs. \eq{mod5}-\eq{mod4}. As a consequence,
the evaluation of \eq{mod5} and \eq{mod4} at $Z=0$, that is
\be \widehat{F'}_{\mu\nu}|_{Z=0}~=~0\ ,\qquad \widehat
D_\mu\widehat\Phi'|_{Z=0}~=~0\ ,\label{final} \ee
give a set of integrable constraints on spacetime that describe a
perturbative expansion of the full field equations.

The main difference between the minimal and modified schemes is
that the latter involves additional parametric integrals
associated with representations of distributional projectors. This
might result in undesirable negative powers of $K^2$ (see
\eq{PP}). Therefore, we are presently inclined to believe that the
minimal scheme will be the one that ultimately leads to a finite
curvature expansion of the $D$-dimensional Vasiliev equations with
supplementary $Sp(2,R)$ invariance conditions.

%%%%%%%%%%%%%%%%%%%%%%%%%%%%%%%%%%%%%%%%%%%%%%%%%%%%%%%%%%%%%%

\section*{Acknowledgments}

%%%%%%%%%%%%%%%%%%%%%%%%%%%%%%%%%%%%%%%%%%%%%%%%%%%%%%%%%%%%%%

We are grateful to D. Anselmi, M. Bianchi, G. Bonelli, A.
Bredthauer, L. Cornalba, U. Danielsson, D. Francia, P. Fr\'e, C.M.
Hull, C. Iazeolla, U. Lindstr\"om, J. Minahan, P. Rajan, L.
Rastelli, B. Sundborg, K. Zarembo, and especially to M. Vasiliev,
for useful discussions. P.S. also wishes to acknowledge the
stimulating collaboration with J. Engquist on the results referred
to in Section 2. The work of A.S. was supported in part by INFN,
by the MIUR-COFIN contract 2003-023852, by the EU contracts
MRTN-CT-2004-503369 and MRTN-CT-2004-512194, by the INTAS contract
03-51-6346, and by the NATO grant PST.CLG.978785. The work of E.S.
was supported in part by the NSF grant PHY-0314712. The work of
P.S. was supported in part by the VR grants 621-2001-1602 and
629-2002-8333. We all benefitted from mutual visits and would like
to thank our Institutions for their kind hospitality and support.

\newpage

%%%%%%%%%%%%%%%%%%%%%%%%%%%%%%%%%%%%%%%%%%%%%%%%%%%%%%%%%%%%%%


\begin{thebibliography}{0}

%%%%%%%%%%%%%%%%%%%%%%%%%%%%%%%%%%%%%%%%%%%%%%%%%%%%%%%%%%%%%


\bibitem{v1} M.~A. Vasiliev,
%``Consistent Equation For Interacting Gauge Fields Of All Spins In
%(3+1)-Dimensions,''
Phys. Lett. {\bf B243} 378 (1990).

\bibitem{bsst} E. Bergshoeff, A. Salam, E. Sezgin and  Y. Tanii,
%{\it Singletons, higher spin massless states and the supermembrane},
Phys. Lett. {\bf 205B}, 237 (1988).

\bibitem{Vasiliev:1999ba} M.~A.~Vasiliev,
%``Higher-spin gauge theories in four, three and two dimensions,''
Int.\ J.\ Mod.\ Phys.\ D5 (1996) 763 [arXiv:hep-th/9611024],
%%CITATION = HEP-TH 9611024;%%
%``Higher spin gauge theories: Star-product and AdS space,''
arXiv:hep-th/9910096;
%%CITATION = HEP-TH 9910096;%%
%``Progress in higher spin gauge theories,''
arXiv:hep-th/0104246.
%%CITATION = HEP-TH 0104246;%%

\bibitem{5d} E. Sezgin and  P. Sundell,
%{\it Doubletons and 5D higher spin gauge theory},
%JHEP {\bf 0109}, 36 (2001),
JHEP {\bf 0109} 036 (2001) [arXiv:hep-th/0105001].

\bibitem{Bouatta:2004kk}
N.~Bouatta, G.~Compere and A.~Sagnotti,
%``An introduction to free higher-spin fields,''
arXiv:hep-th/0409068.
%%CITATION = HEP-TH 0409068;%%

\bibitem{Sundborg:2000wp} B. Sundborg,
%``Stringy gravity, interacting tensionless strings and massless higher
%spins,''
Nucl.\ Phys.\ Proc.\ Suppl.\  {\bf 102}, 113 (2001)
[arXiv:hep-th/0103247]; E. Witten, talk at JHS 60 (Caltech,
November 2001).

\bibitem{Sezgin:2002rt} E. Sezgin and P. Sundell,
%``Massless higher spins and holography,''
Nucl.\ Phys.\ B {\bf 644}, 303 (2002) [Erratum-ibid.\ B {\bf 660},
403 (2003)] [arXiv:hep-th/0205131].

\bibitem{Bianchi:2003wx}
M.~Bianchi, J.F.~Morales and H.~Samtleben,
%``On stringy AdS(5) x S**5 and higher spin holography,''
JHEP {\bf 0307}, 062 (2003) [arXiv:hep-th/0305052]; N.~Beisert,
M.~Bianchi, J.F.~Morales and H.~Samtleben,
%``On the spectrum of AdS/CFT beyond supergravity,''
JHEP {\bf 0402}, 001 (2004) [arXiv:hep-th/0310292]. For a review,
see: M.~Bianchi,
%``Higher spins and stringy AdS(5) x S(5),''
arXiv:hep-th/0409304.
%%CITATION = HEP-TH 0409304;%%

\bibitem{Sezgin:2001yf}
E.~Sezgin and P.~Sundell,
%``Towards massless higher spin extension of D = 5, N = 8 gauged
%supergravity,''
JHEP {\bf 0109}, 025 (2001) [arXiv:hep-th/0107186].

\bibitem{Alkalaev:2002rq}
K.~B.~Alkalaev and M.~A.~Vasiliev,
%``N = 1 supersymmetric theory of higher spin gauge fields in
%AdS(5) at  the cubic level,''
Nucl.\ Phys.\ B {\bf 655}, 57 (2003) [arXiv:hep-th/0206068].

\bibitem{vd} M.~A. Vasiliev,
%{\it Nonlinear equations for symmetric massless higher
% spin fields in $(A)dS_d$},
Phys. Lett. {\bf B567}, 139(2003) [arXiv:hep-th/0304049].

\bibitem{Bars:2001um}
I.~Bars and C.~Kounnas,
%``Theories with two times,''
Phys.\ Lett.\ B {\bf 402} (1997) 25 [arXiv:hep-th/9703060];
%%CITATION = HEP-TH 9703060;%%
I.~Bars and C.~Deliduman,
%``High spin gauge fields and two-time physics,''
Phys.\ Rev.\ D {\bf 64} (2001) 045004 [arXiv:hep-th/0103042].
%%CITATION = HEP-TH 0103042;%%
For reviews see: I.~Bars,
%``Survey of two-time physics,''
Class.\ Quant.\ Grav.\  {\bf 18} (2001) 3113
[arXiv:hep-th/0008164],
%%CITATION = HEP-TH 0008164;%%
%``2T physics 2001,''
arXiv:hep-th/0106021,
%%CITATION = HEP-TH 0106021;%%
%``MSFT: Moyal star formulation of string field theory,''
arXiv:hep-th/0211238.
%%CITATION = HEP-TH 0211238;%%

\bibitem{gunaydin}
M.~Gunaydin and C.~Saclioglu,
%``Oscillator Like Unitary Representations Of Noncompact Groups With A Jordan
%Structure And The Noncompact Groups Of Supergravity,''
Commun.\ Math.\ Phys.\  {\bf 87} (1982) 159;
%%CITATION = CMPHA,87,159;%%
M.~Gunaydin and N.~Marcus,
%``The Spectrum Of The S**5 Compactification Of The Chiral N=2, D = 10
%Supergravity And The Unitary Supermultiplets Of U(2, 2/4),''
Class.\ Quant.\ Grav.\  {\bf 2} (1985) L11;
%%CITATION = CQGRD,2,L11;%%
M.~Gunaydin, D.~Minic and M.~Zagermann,
%``4D doubleton conformal theories, CPT and II B string on AdS(5) x S(5),''
Nucl.\ Phys.\ B {\bf 534} (1998) 96 [Erratum-ibid.\ B {\bf 538}
(1999) 531] [arXiv:hep-th/9806042].
%%CITATION = HEP-TH 9806042;%%

\bibitem{chanpaton}
J.~E.~Paton and H.~M.~Chan,
%``Generalized Veneziano Model With Isospin,''
Nucl.\ Phys.\ B {\bf 10} (1969) 516;
%%CITATION = NUPHA,B10,516;%%
J.~H.~Schwarz,
%``Gauge Groups For Type I Superstrings,''
CALT-68-906-REV
%\href{http://www.slac.stanford.edu/spires/find/hep/www?r=
%calt-68-906-rev}{SPIRES entry}
{\it Presented at 6th Johns Hopkins Workshop on Current Problems
in High-Energy Particle Theory, Florence, Italy, Jun 2-4, 1982};
N.~Marcus and A.~Sagnotti,
%``Tree Level Constraints On Gauge Groups For Type I Superstrings,''
Phys.\ Lett.\ B {\bf 119} (1982) 97,
%%CITATION = PHLTA,B119,97;%%
%``Group Theory From 'Quarks' At The Ends Of Strings,''
Phys.\ Lett.\ B {\bf 188} (1987) 58.
%%CITATION = PHLTA,B188,58;%%
For reviews, see: J.~H.~Schwarz,
%``Superstring Theory,''
Phys.\ Rept.\  {\bf 89} (1982) 223;
%%CITATION = PRPLC,89,223;%%
C.~Angelantonj and A.~Sagnotti,
%``Open strings,''
Phys.\ Rept.\  {\bf 371} (2002) 1 [Erratum-ibid.\  {\bf 376}
(2003) 339] [arXiv:hep-th/0204089].
%%CITATION = HEP-TH 0204089;%%

\bibitem{ps} J. Engquist and P. Sundell, in preparation.

\bibitem{unfolded} M.~A.~Vasiliev,
%``Consistent Equations For Interacting Massless Fields Of
%All Spins In The First Order In Curvatures,''
Annals Phys.\ {\bf 90} (1989) 59;
%%CITATION = APNYA,190,59;%%
%``Unfolded representation for relativistic equations in (2+1)
%anti-De Sitter space,''
Clas.\ Quant.\ Grav.\ {\bf 11} (1994) 649.
%%CITATION = CQGRD,11,649;%%

\bibitem{s3} A. Sagnotti, E. Sezgin and P. Sundell, in preparation.

\bibitem{Bars:2001ma}
I.~Bars,
%``Map of Witten's * to Moyal's *,''
Phys.\ Lett.\ B {\bf 517} (2001) 436 [arXiv:hep-th/0106157];
%%CITATION = HEP-TH 0106157;%%
I.~Bars and S.~J.~Rey,
%``Noncommutative Sp(2,R) gauge theories: A field theory approach to two-time
%physics,''
Phys.\ Rev.\ D {\bf 64} (2001) 046005 [arXiv:hep-th/0104135];
%%CITATION = HEP-TH 0104135;%%%
I.~Bars, I.~Kishimoto and Y.~Matsuo,
%``String amplitudes from Moyal string field theory,''
Phys.\ Rev.\ D {\bf 67} (2003) 066002 [arXiv:hep-th/0211131].
%%CITATION = HEP-TH 0211131;%%

\bibitem{7d} E. Sezgin and P. Sundell,
%{\it 7D bosonic higher spin theory: symmetry algebra
%and linearized constraints},
Nucl. Phys. {\bf B634}, 120 (2002) [arXiv:hep-th/0112100].

\bibitem{mishacubic}
M.~A. Vasiliev,
%``Cubic interactions of bosonic higher spin gauge fields in AdS(5),''
Nucl.\ Phys.\ B {\bf 616}, 106 (2001) [Erratum-ibid.\ B {\bf 652},
407 (2003)] [arXiv:hep-th/0106200].
%%CITATION = HEP-TH 0106200;%%

\bibitem{Francia:2002aa}
D. Francia and A. Sagnotti,
%``Free geometric equations for higher spins,''
Phys.\ Lett.\ B {\bf 543}, 303 (2002) [arXiv:hep-th/0207002],
%``On the geometry of higher-spin gauge fields,''
Class.\ Quant.\ Grav.\  {\bf 20}, S473 (2003)
[arXiv:hep-th/0212185].

\bibitem{Sagnotti:2003qa}
A.~Sagnotti and M.~Tsulaia,
%``On higher spins and the tensionless limit of string theory,''
Nucl.\ Phys.\ B {\bf 682}, 83 (2004) [arXiv:hep-th/0311257].
%%CITATION = HEP-TH 0311257;%%

\bibitem{fronsdal}
C.~Fronsdal,
%``Massless Fields With Integer Spin,''
Phys.\ Rev.\ D {\bf 18}, 3624 (1978).
%%CITATION = PHRVA,D18,3624;%%

\bibitem{bb}
X.~Bekaert and N.~Boulanger,
%``On geometric equations and duality for free higher spins,''
Phys.\ Lett.\ B {\bf 561} (2003) 183 [arXiv:hep-th/0301243],
%``Mixed symmetry gauge fields in a flat background,''
arXiv:hep-th/0310209.
%%CITATION = HEP-TH 0310209;%%

\bibitem{fdw}
B.~de Wit and D.~Z.~Freedman
%``Systematics Of Higher Spin Gauge Fields,''
Phys.\ Rev.\ D {\bf 21} (1980) 358.
%%CITATION = PHRVA,D21,358;%%

\bibitem{susy}
J.~Engquist, E.~Sezgin and P.~Sundell,
%``Superspace formulation of 4D higher spin gauge theory,''
Nucl.\ Phys.\ B {\bf 664}, 439 (2003) [arXiv:hep-th/0211113].
%%CITATION = HEP-TH 0211113;%%

\bibitem{oldlindstrom}
A.~Schild,
%``Classical Null Strings,''
Phys.\ Rev.\ D {\bf 16} (1977) 1722;
%%CITATION = PHRVA,D16,1722;%%
A.~Karlhede and U.~Lindstrom,
%``The Classical Bosonic String In The Zero Tension Limit,''
Class.\ Quant.\ Grav.\  {\bf 3}, (1986) L73;
        %%CITATION = CQGRD,3,L73;%%
H.~Gustafsson, U.~Lindstrom, P.~Saltsidis, B.~Sundborg and R.~von
Unge,
%``Hamiltonian BRST quantization of the conformal string,''
Nucl.\ Phys.\ B {\bf 440}, (1995) 495; [arXiv:hep-th/9410143].
%%CITATION = HEP-TH 9410143;%%
J.~Isberg, U.~Lindstrom, B.~Sundborg, G.~Theodoridis.
%`` Classical and quantized tensionless strings,''
Nucl. \ Phys.\ B {\bf 411} (1994) 122. [arXiv:hep-th/9307108]
%%CITATION = HEP-TH 9307108;%%

\bibitem{pashnev}
M. Henneaux and C. Teitelboim, in ``Quantum Mechanics of
Fundamental Systems, 2'', eds. C. Teitelboim and J. Zanelli
(Plenum Press, New York, 1988), p. 113; A.~Pashnev and
M.~M.~Tsulaia,
%``Dimensional reduction and BRST approach to the description of a Regge trajectory,''
Mod.\ Phys.\ Lett.\ A {\bf 12} (1997) 861 [arXiv:hep-th/9703010];
%%CITATION = HEP-TH 9703010;%%
C.~Burdik, A.~Pashnev and M.~Tsulaia,
%``The Lagrangian description of representations of the Poincare group,''
Nucl.\ Phys.\ Proc.\ Suppl.\  {\bf 102} (2001) 285
[arXiv:hep-th/0103143],
%%CITATION = HEP-TH 0103143;%%
%``Description of the higher massless irreducible integer spins in the  BRST approach,''
Mod.\ Phys.\ Lett.\ A {\bf 13} (1998) 1853 [arXiv:hep-th/9803207];
%%CITATION = HEP-TH 9803207;%%
I.~L.~Buchbinder, V.~A.~Krykhtin and V.~D.~Pershin,
%``Consistent equations for massive spin-2 field coupled to gravity in  string theory,''
Phys.\ Lett.\ B {\bf 466} (1999) 216, {\tt hep-th/9908028};
%%CITATION = HEP-TH 9908028;%%
I.~L.~Buchbinder, D.~M.~Gitman, V.~A.~Krykhtin and V.~D.~Pershin,
%``Equations of motion for massive spin 2 field coupled to gravity,''
Nucl.\ Phys.\ B {\bf 584} (2000) 615, {\tt hep-th/9910188};
%%CITATION = HEP-TH 9910188;%%
I.~L.~Buchbinder, D.~M.~Gitman and V.~D.~Pershin,
%``Causality of massive spin 2 field in external gravity,''
Phys.\ Lett.\ B {\bf 492} (2000) 161, {\tt hep-th/0006144};
%%CITATION = HEP-TH 0006144;%%
I.~L.~Buchbinder and V.~D.~Pershin,
% ``Gravitational interaction of higher spin massive fields and string  theory''
{\tt hep-th/0009026};
%%CITATION = HEP-TH 0009026;%%
I.~L.~Buchbinder, V.~A.~Krykhtin and A.~Pashnev,
%``BRST approach to Lagrangian construction for fermionic massless higher spin
%fields,''
arXiv:hep-th/0410215.
%%CITATION = HEP-TH 0410215;%%

\bibitem{thorn}
C.~B.~Thorn,
%``Reformulating string theory with the 1/N expansion,''
arXiv:hep-th/9405069,
%%CITATION = HEP-TH 9405069;%%
%``Calculating the rest tension for a polymer of string bits,''
Phys.\ Rev.\ D {\bf 51} (1995) 647 [arXiv:hep-th/9407169];
%%CITATION = HEP-TH 9407169;%%
O.~Bergman and C.~B.~Thorn,
%``String bit models for superstring,''
Phys.\ Rev.\ D {\bf 52} (1995) 5980 [arXiv:hep-th/9506125].
%%CITATION = HEP-TH 9506125;%%
For a review see: O.~Bergman,
%``Bits of string and bits of branes,''
arXiv:hep-th/9607183.
%%CITATION = HEP-TH 9607183;%%

\bibitem{deser}
S.~Deser and A.~Waldron,
%``Gauge invariances and phases of massive higher spins in (A)dS,''
Phys.\ Rev.\ Lett.\  {\bf 87} (2001) 031601 {\tt hep-th/0102166};
%%CITATION = HEP-TH 0102166;%%
%``Partial masslessness of higher spins in (A)dS,''
Nucl.\ Phys.\ B {\bf 607} (2001) 577 {\tt hep-th/0103198};
%%CITATION = HEP-TH 0103198;%%
%``Stability of massive cosmological gravitons,''
Phys.\ Lett.\ B {\bf 508} (2001) 347 {\tt hep-th/0103255};
%%CITATION = HEP-TH 0103255;%%
%``Null propagation of partially massless higher spins in (A)dS and cosmological constant speculations,''
Phys.\ Lett.\ B {\bf 513} (2001) 137 {\tt hep-th/0105181}.
%%CITATION = HEP-TH 0105181;%%

\bibitem{petkou}
R.~R.~Metsaev,
%``Massless arbitrary spin fields in AdS(5),''
Phys.\ Lett.\ B {\bf 531} (2002) 152 [arXiv:hep-th/0201226];
%%CITATION = HEP-TH 0201226;%%
I.~R.~Klebanov and A.~M.~Polyakov,
%``AdS dual of the critical O(N) vector model,''
Phys.\ Lett.\ B {\bf 550} (2002) 213 [arXiv:hep-th/0210114];
%%CITATION = HEP-TH 0210114;%%
L.~Girardello, M.~Porrati and A.~Zaffaroni,
%``3-D interacting CFTs
%and generalized Higgs phenomenon in higher spin theories on AdS,''
Phys.\ Lett.\ B {\bf 561} (2003) 289 [arXiv:hep-th/0212181];
%%CITATION = HEP-TH 0212181;%%
R.~G.~Leigh and A.~C.~Petkou,
%``Holography of the N = 1 higher-spin theory on AdS(4),''
JHEP {\bf 0306} (2003) 011 [arXiv:hep-th/0304217],
%%CITATION = HEP-TH 0304217;%%
%``SL(2,Z) action on three-dimensional CFTs and holography,''
arXiv:hep-th/0309177; H.~J.~Schnitzer,
%``Gauged vector models and higher-spin representations in AdS(5),''
arXiv:hep-th/0310210.
%%CITATION = HEP-TH 0310210;%%

\bibitem{Lindstrom:2003mg}
U.~Lindstrom and M.~Zabzine,
%``Tensionless strings, WZW models at critical level and massless higher  spin
%fields,''
Phys.\ Lett.\ B {\bf 584} (2004) 178 [arXiv:hep-th/0305098].
%%CITATION = HEP-TH 0305098;%%
G.~Bonelli,
%``On the tensionless limit of bosonic strings, infinite symmetries and  higher spins,''
Nucl.\ Phys.\ B {\bf 669} (2003) 159 [arXiv:hep-th/0305155],
%%CITATION = HEP-TH 0305155;%%
%``On the covariant quantization of tensionless bosonic strings in AdS spacetime,''
JHEP {\bf 0311} (2003) 028 [arXiv:hep-th/0309222].
%%CITATION = HEP-TH 0309222;%%

\bibitem{Bakas:2004jq}
M.~Plyushchay, D.~Sorokin and M.~Tsulaia,
%``Higher spins from tensorial charges and OSp(N$|$2n) symmetry,''
JHEP {\bf 0304} (2003) 013 [arXiv:hep-th/0301067];
%%CITATION = HEP-TH 0301067;%%
G.~K.~Savvidy,
%``Tensionless strings: Physical Fock space and higher spin fields,''
Int.\ J.\ Mod.\ Phys.\ A {\bf 19} (2004) 3171
[arXiv:hep-th/0310085];
%%CITATION = HEP-TH 0310085;%%
I.~Bakas and C.~Sourdis,
%``On the tensionless limit of gauged WZW models,''
JHEP {\bf 0406} (2004) 049 [arXiv:hep-th/0403165];
%%CITATION = HEP-TH 0403165;%%
J.~Mourad,
%``Continuous spin and tensionless strings,''
arXiv:hep-th/0410009.
%%CITATION = HEP-TH 0410009;%%

\bibitem{misha}
M.~A.~Vasiliev,
%``Higher spin superalgebras in any dimension and their representations,''
arXiv:hep-th/0404124.
%%CITATION = HEP-TH 0404124;%%

\bibitem{Gunaydin:1989um}
M.~Gunaydin,
%``Singleton And Doubleton Supermultiplets Of
%Space-Time Supergroups And Infinite Spin Superalgebras,''
CERN-TH-5500/89, {\it Invited talk given at Trieste Conf. on
Supermembranes and Physics in (2+1)-Dimensions, Trieste, Italy,
Jul 17-21, 1989}

%\cite{eastwood}
\bibitem{eastwood}
M.~G.~Eastwood,
%``Higher symmetries of the Laplacian,''
arXiv:hep-th/0206233.
%%CITATION = HEP-TH 0206233;%%

\bibitem{Flato:1978qz}
M.~Flato and C.~Fronsdal,
%``One Massless Particle Equals Two Dirac Singletons: Elementary Particles In A
%Curved Space. 6,''
Lett.\ Math.\ Phys.\  {\bf 2} (1978) 421.
%%CITATION = LMPHD,2,421;%%

\bibitem{mishaframe}
M.~A.~Vasiliev,
%``Free Massless Fields Of Arbitrary Spin In The De Sitter Space And Initial
%Data For A Higher Spin Superalgebra,''
Fortsch.\ Phys.\  {\bf 35} (1987) 741 [Yad.\ Fiz.\  {\bf 45}
(1987) 1784];
%%CITATION = FPYKA,35,741;%%
V.~E.~Lopatin and M.~A.~Vasiliev,
%``Free Massless Bosonic Fields Of Arbitrary Spin In D-Dimensional De Sitter
%Space,''
Mod.\ Phys.\ Lett.\ A {\bf 3} (1988) 257.
%%CITATION = MPLAE,A3,257;%%

\bibitem{CF}
M.~Kontsevich,
%``Deformation quantization of Poisson manifolds, I,''
Lett.\ Math.\ Phys.\  {\bf 66} (2003) 157 [arXiv:q-alg/9709040];
%%CITATION = Q-ALG 9709040;%%
A.~S.~Cattaneo and G.~Felder,
%``On the globalization of Kontsevich's star product and the perturbative
%Poisson sigma model,''
Prog.\ Theor.\ Phys.\ Suppl.\  {\bf 144} (2001) 38
[arXiv:hep-th/0111028].
%%CITATION = HEP-TH 0111028;%%

\bibitem{dfre} R.~D'Auria and P.~Fr\'e,
%``Geometric Supergravity In D = 11 And Its Hidden Supergroup,''
Nucl.\ Phys.\ B {\bf 201} (1982) 101 [Erratum-ibid.\ B {\bf 206}
(1982) 496];
%%CITATION = NUPHA,B201,101;%%
L.~Castellani, P.~Fr\'e, F.~Giani, K.~Pilch and P.~van
Nieuwenhuizen,
%``Beyond D = 11 Supergravity And Cartan Integrable Systems,''
Phys.\ Rev.\ D {\bf 26} (1982) 1481;
%%CITATION = PHRVA,D26,1481;%%
R.~D'Auria and P.~Fr\'e,
 %``Cartan Integrable Systems, That Is Differential Free Algebras, In
%Supergravity,''
Print-83-0689 (TURIN)
%\href{http://www.slac.stanford.edu/spires/find/hep/www?r=
%print-83-0689\%2F(turin)}{SPIRES entry}
{\it Lectures given at the September School on Supergravity and
Supersymmetry, Trieste, Italy, Sep 6-18, 1982}; P.~Fr\'e,
%``Comments On The Six Index Photon In D = 11 Supergravity
%And The Gauging Of Free Differential Algebras,''
Class.\ Quant.\ Grav.\  {\bf 1} (1984)L81.
%%CITATION = CQGRD,1,L81;%%

\bibitem{Kristiansson:2003xx}
F. Kristiansson and P. Rajan,
%``Scalar field corrections to AdS(4) gravity from higher spin gauge  theory,''
JHEP {\bf 0304}, 009 (2003) [arXiv:hep-th/0303202].

\bibitem{Sezgin:2003pt} E. Sezgin and P. Sundell,
%``Holography in 4D (super) higher spin theories and a test
%via cubic  scalar couplings,''
arXiv:hep-th/0305040.

%\cite{misha2}
\bibitem{misha2}
O.~V.~Shaynkman and M.~A.~Vasiliev,
%``Scalar field in any dimension from the higher spin gauge theory
%perspective,''
Theor.\ Math.\ Phys.\  {\bf 123} (2000) 683 [Teor.\ Mat.\ Fiz.\
{\bf 123} (2000) 323] [arXiv:hep-th/0003123].
%%CITATION = HEP-TH 0003123;%%

\bibitem{dvh}
M.~Dubois-Violette and M.~Henneaux,
%``Generalized cohomology for irreducible tensor fields of mixed Young
%symmetry type,'' Lett.\ Math.\ Phys.\  {\bf 49} (1999) 245
[arXiv:math.qa/9907135],
%%CITATION = MATH-QA 9907135;%%
%``Tensor fields of mixed Young symmetry type and N-complexes,''
Commun.\ Math.\ Phys.\  {\bf 226} (2002) 393
[arXiv:math.qa/0110088].
%%CITATION = MATH-QA 0110088;%%

\bibitem{ww}
See, for instance, E.~Whittaker and G.N.~Watson, ``A Course of
Modern Analysis'' (Cambridge, Cambridge Univ. Press, 1996).

%\cite{Salam:1981xd}
\bibitem{Salam:1981xd}
A.~Salam and J.~Strathdee,
%``On Kaluza-Klein Theory,''
Annals Phys.\  {\bf 141} (1982) 316.
%%CITATION = APNYA,141,316;%%
\end{thebibliography}
\end{document}